\documentclass[11pt,a4paper]{article}
\pdfoutput=1
\usepackage[T1]{fontenc}
\usepackage{jheppub}
\usepackage{mathrsfs}
\usepackage{esint}
\usepackage{xspace}
\input{notation.tex}
\graphicspath{{figs/}}

\title{Holographic two-point functions across a PT symmetry-breaking transition}

\author[a,b,c]{Zuo Wang}
\author[a,b,c]{and Liang He}
\affiliation[a]{Institute for Theoretical Physics, School of Physics, South China Normal University, Guangzhou 510006, China}
\affiliation[b]{Key Laboratory of Atomic and Subatomic Structure and Quantum Control (Ministry of Education), Guangdong Basic Research Center of Excellence for Structure and Fundamental Interactions of Matter, School of Physics, South China Normal University, Guangzhou 510006, China}
\affiliation[c]{Guangdong Provincial Key Laboratory of Quantum Engineering and Quantum Materials, Guangdong-Hong Kong Joint Laboratory of Quantum Matter, South China Normal University, Guangzhou 510006, China}
\emailAdd{liang.he@scnu.edu.cn}

\abstract{
In parity-time (PT)-symmetric non-Hermitian systems, an exceptional point, where eigenvalues and eigenvectors coalesce, typically separates a real spectrum from complex-conjugate eigenvalues.
In logarithmic conformal field theories (LCFTs), such an exceptional point appears in the spectrum of scaling dimensions.
Holographic models of LCFTs, such as the dipole model, are built exactly at this point.
Here, we deform the holographic dipole model so that its two coalescing scaling dimensions split into either two real ones or a complex-conjugate pair.
We obtain the renormalized two-point functions in closed form.
They change continuously from a linear combination of two power laws, through the logarithmic form of the LCFT at the exceptional point, to log-periodic oscillations.
For any pair of scalar primary operators with real mixing under dilatations, the conformal Ward identities fix this form up to two normalization constants, which the holographic dictionary determines.
In PT-symmetric theories with such a dilatation operator, real and complex-conjugate scaling dimensions correspond to the PT-unbroken and PT-broken phases.
In the BTZ black hole at finite temperature, the difference between the quasinormal frequencies associated with the two scaling dimensions changes from purely imaginary to purely real across the real-to-complex spectral transition. 
Correspondingly, the linear response changes from overdamped to underdamped, with critical damping at the exceptional point.
Direct experimental observation of the two-point functions across the exceptional point could be achieved in two hyperbolic electric circuits with a non-reciprocal coupling, through the voltage response between their edge nodes.
}

\begin{document}
\maketitle \flushbottom

\section{Introduction}

In logarithmic conformal field theories (LCFTs), the dilatation operator is non-diagonalizable and contains Jordan blocks \cite{GURARIE1993NPB,Hogervorst2017JHEP}.
LCFTs describe critical behavior in percolation, polymers, and systems with quenched disorder \cite{CAUX1996NPB,Cardy2013JPA}.
Through the AdS/CFT correspondence \cite{Maldacena1999IJTP,witten1998ATMP,GUBSER1998PLB}, higher-derivative theories in anti-de Sitter (AdS) spacetime provide holographic realizations of LCFTs \cite{Grumiller2013JPA,Hogervorst2017JHEP}.
Examples include the scalar dipole model \cite{GHEZELBASH1999IJMPA,KOGAN1999PLB}, originally developed in the study of singletons \cite{Flato1987CMP} and generalized to LCFTs of higher rank \cite{Bergshoeff2012JHEP}.
Another example is critical gravity \cite{Lu2011PRL,Bergshoeff2011PRD}, in which the stress tensor and its logarithmic partner form a rank-2 Jordan block \cite{Grumiller2013JPA,Johansson2012JHEP}.
A related Jordan-block structure appears in the parity-time (PT)-symmetric quantization of conformal gravity linearized about flat spacetime \cite{BenderMannheim2008PRD,Mannheim2011GRG,Mannheim2023EPJP}.
More recently, Jordan-block structures were also proposed in flat-space holography, for the dilatation operator in celestial conformal field theory \cite{Fiorucci2024PRD,Bissi2024JHEP,Bhardwaj2024JHEP} and for the boost generators in Carrollian field theories \cite{Hao2022PRD,Bagchi2026EPJC}.

The Jordan blocks of an LCFT are directly visible in its two-point functions \cite{GURARIE1993NPB,Hogervorst2017JHEP}.
For a rank-2 block, the correlator of the eigenoperator with itself vanishes, whereas that of its logarithmic partner contains a logarithm of the distance. 
Such logarithms can arise when two real scaling dimensions and the corresponding operators coalesce \cite{Cardy2013JPA,Hogervorst2017JHEP}.
In critical percolation, for instance, the coalescence of two operators of the $Q$-state Potts model as $Q\rightarrow1$ produces a logarithm with a universal prefactor \cite{Vasseur2012JSTAT}.
Holographic calculations reproduce these correlators in the scalar dipole model \cite{GHEZELBASH1999IJMPA,KOGAN1999PLB,Bergshoeff2012JHEP} and in critical gravity \cite{Johansson2012JHEP}.
In critical gravity, the stress tensor has a vanishing two-point function with itself but a nonzero one with its logarithmic partner \cite{Johansson2012JHEP}.
In two-dimensional LCFTs, the coefficient of the two-point function between the stress tensor and its logarithmic partner plays a role analogous to the central charge \cite{Gurarie1999NPB,Skenderis2009JHEP}.
In both percolation and gravity duals of LCFTs, the coefficient of the logarithm is set by the normalization of the two-point functions relative to the splitting of the two real dimensions before they coalesce \cite{Skenderis2009JHEP,Grumiller2011JHEP,Vasseur2012JSTAT}.
Two-point functions thus carry information beyond the scaling dimensions themselves.
Moreover, in these examples, the two scaling dimensions cross at the coalescence point and remain real on both sides \cite{Skenderis2009JHEP,Vasseur2012JSTAT}.

From the perspective of non-Hermitian physics, the coalescence point of an LCFT is an exceptional point of the dilatation operator, which plays the role of the Hamiltonian in radial quantization \cite{Grumiller2013JPA,Hogervorst2017JHEP,Ashida2020AP}.
Near a generic exceptional point of a real matrix, by contrast, the two eigenvalues split as the square root of the deformation, from a real pair on one side to a complex-conjugate pair on the other \cite{Ashida2020AP}.
In PT-symmetric systems, such an exceptional point separates the PT-unbroken phase with real eigenvalues from the PT-broken phase with complex-conjugate eigenvalues \cite{Ashida2020AP,bender2023arXiv}.
Recently, exceptional points of the dilatation operator were identified in perturbative Yang--Mills theory \cite{Jin2026arXiv} and at conformal interfaces between two-dimensional rational CFTs \cite{Chen2026arXiv}.
Complex scaling dimensions also arise when two fixed points merge and annihilate \cite{Kaplan2009PRD,Gorbenko2018JHEP,Faedo2020PRL}.
In holography, PT symmetry-breaking transitions are studied with non-Hermitian boundary sources \cite{Daniel_2020_SciPost,Sergio_2023_SciPost,Meyer_2024_SciPost}. 
PT-symmetric holographic duals also arise from supergravity solutions with imaginary scalars \cite{Begines2026arXiv}.
In the BTZ black hole, mixed boundary conditions can also produce exceptional points in the quasinormal-mode spectrum, where two poles of the retarded two-point function coalesce into a second-order pole \cite{WangYang2026JHEP,Long2026arXiv}.
The linear response here then acquires a term linear in time \cite{WangYang2026JHEP,Long2026arXiv}.
A natural question that arises in this context is how holographic two-point functions behave when the scaling dimensions of an LCFT are continued through the exceptional point into a complex-conjugate pair.
A related question is how these two-point functions describe the PT-unbroken and PT-broken phases of theories whose dilatation operator mixes two scalar primary operators in the same way.
Electric circuits offer a laboratory platform for both holography on hyperbolic lattices \cite{Lenggenhager2022NC,Chen2023arXiv,Dey2024PRL} and non-Hermitian physics \cite{Schindler2011PRA,Helbig2020NP,Stegmaier2021PRL}.
This raises the question whether the two-point functions across the exceptional point can be observed experimentally in such circuits.

In this work, we deform the dipole model in Euclidean AdS$_{d+1}$ by one parameter, so that the two coalescing scaling dimensions split into either two real ones or a complex-conjugate pair [see eq.~(\ref{eq:the_action}) and figure~\ref{fig:dimensions}].
We find the following.
(i) Two-point functions across the exceptional point.
The renormalized holographic two-point functions are linear combinations of two power laws when the two scaling dimensions are real and exhibit log-periodic oscillations when they are complex conjugate [see eqs.~(\ref{eq:PT_AdS_2pt_1})--(\ref{eq:PT_AdS_2pt_3}) and figure~\ref{fig:two_point}].
At the exceptional point, they continuously reduce to those of the LCFT dual to the dipole model.
In any dimension $d\geq2$, the full conformal Ward identities fix the same form up to two normalization constants, for every real deformation of the exceptional point within a pair of scalar primary operators [see eqs.~(\ref{eq:perturbed_D}) and (\ref{eq: G_solution})].
In PT-symmetric theories with such a dilatation matrix, these two-point functions thus describe the PT-unbroken phase, the PT-broken phase, and the exceptional point separating them (see section~\ref{sec: PT_transition}).
(ii) Real-time signatures of the transition.
In global AdS, the normalizable modes grow and decay in time for complex-conjugate scaling dimensions and acquire a term linear in time at the exceptional point [see eq.~(\ref{eq: global_modes})].
At finite temperature, the linear response in the BTZ black hole ($d=2$) is overdamped for real scaling dimensions, critically damped at the exceptional point, and underdamped for complex-conjugate ones [see eq.~(\ref{eq:thermal_retarded}) and figure~\ref{fig:thermal}].
(iii) Realization in electric circuits.
In two hyperbolic electric circuits with a non-reciprocal coupling, the voltage response between edge nodes directly reflects the two-point functions [see eq.~(\ref{eq:circuit_eps}) and figure~\ref{fig:circuit_measurement}(a)].
At fixed nodes, the voltage varies continuously with the coupling, in agreement with the two-point functions [see figure~\ref{fig:circuit_measurement}(b)].
As a function of the separation, the voltage response follows the two-point functions, including the logarithm of the LCFT (see figure~\ref{fig:circuit}).

This paper is organized as follows.
In section~\ref{sec:model}, we introduce the model and decouple it into two eigenmodes, whose squared masses are real or complex conjugate depending on the sign of the deformation.
In section~\ref{sec:AdS_calculation}, we compute the holographic two-point functions and their limit at the exceptional point.
Section~\ref{sec:CFT_calculation} derives the two-point functions from the conformal Ward identities and matches them with the holographic results.
In section~\ref{sec: PT_transition}, we discuss the conditions under which the spectral transition corresponds to PT symmetry breaking and examine its real-time signatures.
Section~\ref{sec:circuit} realizes the model in two coupled hyperbolic electric circuits.
Section~\ref{sec: Discussion} summarizes the results and gives an outlook.

\section{The model and its eigenmodes}\label{sec:model}

In this section, we introduce a one-parameter deformation of the dipole model in $(d+1)$-dimensional Euclidean AdS spacetime.
We decouple it into two eigenmodes, whose squared masses are real or complex conjugate depending on the sign of the deformation parameter $\epsilon$, and coalesce at $\epsilon=0$.

We consider the action
\begin{equation}
S=-\frac{1}{2}\int d^{d+1}x\sqrt{g}\left[2\nabla_{\mu}\phi_{1}\nabla^{\mu}\phi_{2}+2m^{2}\phi_{1}\phi_{2}+\phi_{1}^{2}+\epsilon\phi_{2}^{2}\right],\label{eq:the_action}
\end{equation}
where $g_{\mu\nu}$ is the fixed AdS metric, $g\equiv\det(g_{\mu\nu})$, $\nabla_{\mu}$ is the covariant derivative, and $\phi_{1}$ and $\phi_{2}$ are real scalar fields.
We take $m^{2}+d^{2}/4>0$ and real $\epsilon<(m^{2}+d^{2}/4)^{2}$.
We treat the bulk fields as classical fields on this fixed background.
Their equations of motion and renormalized on-shell action determine quantum correlation functions of the dual CFT at leading semiclassical order via the holographic prescription.
The action (\ref{eq:the_action}) is motivated by the dipole model \cite{GHEZELBASH1999IJMPA,KOGAN1999PLB}, originally developed in the study of singletons \cite{Flato1987CMP}, which is recovered at $\epsilon=0$,
\begin{equation}
S_{\text{sd}}=-\frac{1}{2}\int d^{d+1}x\sqrt{g}\left[2\nabla_{\mu}\phi_{1}\nabla^{\mu}\phi_{2}+2m^{2}\phi_{1}\phi_{2}+\phi_{1}^{2}\right].\label{eq:dipole_model}
\end{equation}
It is the simplest case of the scalar models in AdS with coupled fields of degenerate masses, whose duals are LCFTs \cite{Bergshoeff2012JHEP}.
The term $\epsilon\phi_{2}^{2}$ in (\ref{eq:the_action}) deforms this model away from the logarithmic point, as we show below.

The equations of motion of the action (\ref{eq:the_action}) are
\begin{equation}
(\nabla_{\mu}\nabla^{\mu}-m^{2})\phi_{1}=\epsilon\phi_{2}, \qquad (\nabla_{\mu}\nabla^{\mu}-m^{2})\phi_{2}=\phi_{1},\label{eq:PT_modified_EOM}
\end{equation}
where the covariant derivatives are given by $\nabla_{\mu}\nabla^{\mu}\phi=z^{d+1}\partial_{\mu}(z^{-d-1}g^{\mu\nu}\partial_{\nu}\phi)$ in \Poincare coordinates ($z,x^{i}$) ($i=1,\dots,d$).
In matrix form, they read
\begin{equation}
(\nabla_{\mu}\nabla^{\mu}-m^{2})\left(\begin{array}{c}
\phi_{1}\\
\phi_{2}
\end{array}\right)=M\left(\begin{array}{c}
\phi_{1}\\
\phi_{2}
\end{array}\right),\quad M\equiv\left(\begin{array}{cc}
0 & \epsilon\\
1 & 0
\end{array}\right).\label{eq:mixing_matrix}
\end{equation}
Eliminating $\phi_{1}$, we can obtain a single fourth-order equation for $\phi_{2}$, $(\nabla_{\mu}\nabla^{\mu}-m^{2})^{2}\phi_{2}=\epsilon\phi_{2}$, or equivalently, 
\begin{equation}
(\nabla_{\mu}\nabla^{\mu}-m^{2}-\sqrt{\epsilon})(\nabla_{\mu}\nabla^{\mu}-m^{2}+\sqrt{\epsilon})\phi_{2}=0.\label{eq:fourth_order}
\end{equation}
This is a field-theory version of a fourth-order oscillator \cite{Pais1950PR,BenderMannheim2008PRL}.
For $\epsilon>0$, eq.~(\ref{eq:fourth_order}) is the fourth-order scalar equation with two different masses of ref.~\cite{Bergshoeff2011JHEP}, whose equal-mass limit gives a logarithmic pair.
Two classical mode frequencies are real and different for $\epsilon>0$, equal for $\epsilon=0$, and complex conjugate for $\epsilon<0$.
For the quantized fourth-order oscillator, equal frequencies give a non-diagonalizable Hamiltonian \cite{BenderMannheim2008PRD}.
Complex-conjugate frequencies instead give complex-conjugate energies \cite{Mannheim2018JPA}.
Up to a field redefinition and an overall sign, the action (\ref{eq:the_action}) is the most general real quadratic action of two scalar fields that admits an exceptional point of its mass matrix (see appendix~\ref{app:generality}).

We can decouple the equations of motion (\ref{eq:PT_modified_EOM}) by introducing two eigenmodes, $\psi_{+}$ and $\psi_{-}$, satisfying $(\nabla_{\mu}\nabla^{\mu}-m^{2})\psi_{\pm}=\lambda_{\pm}\psi_{\pm}$.
Assuming the mixing, $\psi_{\pm}=p\phi_{1}+q\phi_{2}$, we can write
\begin{align}
p\lambda_{\pm}\phi_{1}+q\lambda_{\pm}\phi_{2}=\lambda_{\pm}\psi_{\pm} 
& =(\nabla_{\mu}\nabla^{\mu}-m^{2})\psi_{\pm}=(\nabla_{\mu}\nabla^{\mu}-m^{2})(p\phi_{1}+q\phi_{2})\nonumber \\
& =p\epsilon\phi_{2}+q\phi_{1},
\end{align}
from which we obtain $p\lambda_{\pm}=q$, $q\lambda_{\pm}=p\epsilon$ and hence $\lambda_{\pm}^{2}=\epsilon$.
We choose $\lambda_{\pm}=\pm\sqrt{\epsilon}$, and find two solutions $p=1/2$, $q=\pm\sqrt{\epsilon}/2$.
The mapping that transforms $\phi$ to $\psi$ and its inverse become
\begin{align}
\psi_{\pm} &=\frac{1}{2}\left(\phi_{1}\pm\sqrt{\epsilon}\phi_{2}\right),\\
\phi_{1} &=\psi_{+}+\psi_{-},\qquad\phi_{2}=\frac{1}{\sqrt{\epsilon}}\left(\psi_{+}-\psi_{-}\right).\label{eq:field_map}
\end{align}
The action (\ref{eq:the_action}) transforms as
\begin{align}
S= & -\frac{1}{\sqrt{\epsilon}}\int d^{d+1}x\sqrt{g}\left[\nabla_{\mu}\psi_{+}\nabla^{\mu}\psi_{+}+(m^{2}+\sqrt{\epsilon})\psi_{+}^{2}\right]\nonumber \\
 & +\frac{1}{\sqrt{\epsilon}}\int d^{d+1}x\sqrt{g}\left[\nabla_{\mu}\psi_{-}\nabla^{\mu}\psi_{-}+(m^{2}-\sqrt{\epsilon})\psi_{-}^{2}\right],\label{eq:action_in_psi}
\end{align}
and the equations of motion become
\begin{equation}
(\nabla_{\mu}\nabla^{\mu}-(m^{2}+\sqrt{\epsilon}))\psi_{+}=0,\qquad(\nabla_{\mu}\nabla^{\mu}-(m^{2}-\sqrt{\epsilon}))\psi_{-}=0.\label{eq: classical_splitting}
\end{equation}
Now, we obtain two decoupled modes with squared masses $m^{2}\pm\sqrt{\epsilon}$, which are real for positive $\epsilon$ and complex conjugate for negative $\epsilon$.
For $\epsilon>0$, the two modes in the action (\ref{eq:action_in_psi}) have opposite signs of kinetic terms.
Quantization with the conventional inner product then gives one negative-norm (ghost) mode \cite{BenderMannheim2008PRL,Bergshoeff2011JHEP}.
The dual CFT is therefore non-unitary, as is typical for LCFTs (see also section~\ref{sec: PT_transition}).

Through the AdS/CFT correspondence, the eigenmodes $\psi_{\pm}$ are dual to quantum scalar operators $\mathscr{O}_{\pm}$ with scaling dimensions \cite{witten1998ATMP}
\begin{equation}
\Delta_{\pm}=\frac{d+\sqrt{d^{2}+4\left(m^{2}\pm\sqrt{\epsilon}\right)}}{2},\label{eq:Delta_pm}
\end{equation}
which satisfy $\Delta_{\pm}(\Delta_{\pm}-d)=m^{2}\pm\sqrt{\epsilon}$.
Here, we choose the plus sign before the square root in the AdS mass-dimension relation \cite{Klebanov1999NPB,Faulkner2011JHEP}.
For later convenience, we denote their mean and difference by $\bar{\Delta}\equiv(\Delta_{+}+\Delta_{-})/2$ and $\delta\equiv(\Delta_{+}-\Delta_{-})/2$, and define $\eta\equiv\delta^{2}$.
Subtracting the two relations gives $\sqrt{\epsilon}=\beta\delta$, i.e., 
\begin{equation}
\eta=\frac{\epsilon}{\beta^{2}},\qquad\beta\equiv\Delta_{+}+\Delta_{-}-d=2\bar{\Delta}-d.\label{eq:eta_epsilon}
\end{equation}
As $\epsilon\rightarrow0$, $\Delta_{\pm}\rightarrow\Delta\equiv(d+\sqrt{d^{2}+4m^{2}})/2$ and $\beta\rightarrow\sqrt{d^{2}+4m^{2}}$.

\begin{figure}[t]
\centering
\includegraphics{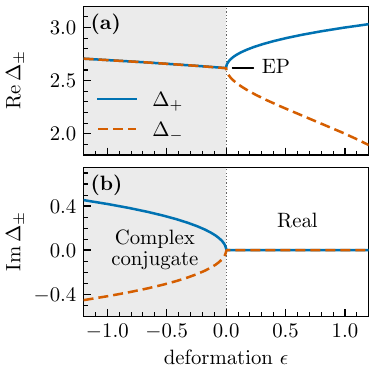}
\caption{Scaling dimensions $\Delta_{\pm}$ as a function of $\epsilon$. \textbf{(a,b)} The real \textbf{(a)} and imaginary \textbf{(b)} parts of $\Delta_{\pm}$, for $d=3$ and $m^{2}=-1$. The two dimensions are real and different for $\epsilon>0$, coalesce at the exceptional point (EP) $\epsilon=0$, and form a complex-conjugate pair for $\epsilon<0$ (shaded). In PT-symmetric theories with the same structure, the two regimes are the PT-unbroken and PT-broken phases (section~\ref{sec: PT_transition}). See section~\ref{sec:model} for details.}
\label{fig:dimensions}
\end{figure}

The sign of $\epsilon$ thus separates three cases (figure~\ref{fig:dimensions}):
\begin{enumerate}
\item[(i)] $0<\epsilon<(m^{2}+d^{2}/4)^{2}$: the two squared masses are real and different, and so are the two scaling dimensions, with $\Delta_{\pm}>d/2$.
\item[(ii)] $\epsilon=0$: the two modes coalesce and the mapping (\ref{eq:field_map}) becomes singular.
The matrix $M$ in (\ref{eq:mixing_matrix}) is a non-diagonalizable rank-2 Jordan block.
This is an exceptional point, where the eigenvalues $\pm\sqrt{\epsilon}$ of $M$ and the corresponding eigenvectors coalesce \cite{Ashida2020AP}.
The model then reduces to the dipole model (\ref{eq:dipole_model}), dual to an LCFT \cite{Bergshoeff2012JHEP}.
\item[(iii)] $\epsilon<0$: $\sqrt{\epsilon}=i\sqrt{|\epsilon|}$, and the squared masses $m^{2}\pm i\sqrt{|\epsilon|}$ are complex conjugate.
With the principal branch of the square root in (\ref{eq:Delta_pm}), the scaling dimensions are complex conjugate, $\Delta_{-}=\Delta_{+}^{*}$, with $\mathrm{Re}\,\Delta_{\pm}>d/2$.
The eigenmodes are also complex conjugate, $\psi_{-}=\psi_{+}^{*}$.
\end{enumerate}
In PT-symmetric theories with the same structure, case (i) is the PT-unbroken phase, case (iii) is the PT-broken phase, and case (ii) is the exceptional point separating them.
In section~\ref{sec: PT_transition}, we discuss under which conditions this identification applies to the present model.

\section{Two-point functions on the AdS side}\label{sec:AdS_calculation}

This section derives the holographic two-point functions of the model (\ref{eq:the_action}).
Since the eigenmodes $\psi_{\pm}$ in (\ref{eq: classical_splitting}) are decoupled fields of squared masses $m^{2}\pm\sqrt{\epsilon}$, we first construct their bulk-to-boundary propagators and the on-shell action, and then map the results to the original fields $\phi_{1}$ and $\phi_{2}$.
The resulting two-point functions are linear combinations of two power laws for $\epsilon>0$, exhibit log-periodic oscillations for $\epsilon<0$, and continuously reduce to those of the LCFT dual to the dipole model as $\epsilon\rightarrow0$ (see figure~\ref{fig:two_point}).

\subsection{Bulk-to-boundary propagators and on-shell action}

We work with \Poincare coordinates ($z,x^{i}$) ($i=1,\dots,d$) for the Euclidean $\mathrm{AdS}_{d+1}$ metric, $ds^{2}=(dz^{2}+dx^{i}dx^{i})/z^{2}$, with the boundary at $z=0$.
Each eigenmode obeys the Klein--Gordon equation (\ref{eq: classical_splitting}), whose bulk-to-boundary propagator is derived in ref.~\cite{witten1998ATMP},
\begin{equation}
\psi_{\pm}(z,x^{i})=\int d^{d}x^{\prime}K_{\Delta_{\pm}}(z,x^{i};0,x^{i\prime})u_{\pm}(x^{i\prime}),
\quad 
K_{\Delta}(z,x^{i};0,x^{i\prime})\equiv\frac{1}{N_{\Delta}}\frac{z^{\Delta}}{(z^{2}+\left|x-x^{\prime}\right|^{2})^{\Delta}},\label{eq:propagator}
\end{equation}
where $\left|x-x^{\prime}\right|^{2}\equiv\sum_{i=1}^{d}(x^{i}-x^{i\prime})^{2}$, $N_{\Delta}\equiv\pi^{d/2}\frac{\Gamma(\Delta-d/2)}{\Gamma(\Delta)}$, and $\Delta_{\pm}$ are given in (\ref{eq:Delta_pm}).
Note that the normalization $1/N_{\Delta}$ is fixed to ensure the boundary behavior $\lim_{z\rightarrow0}\psi_{\pm}(z,x^{i})=\lim_{z\rightarrow0}z^{d-\Delta_{\pm}}u_{\pm}(x^{i})$, using the Dirac delta function \cite{witten1998ATMP,MuckPRD1998} 
\begin{equation}
\delta^{d}(x-x^{\prime})=\lim_{z\rightarrow0}\frac{1}{N_{\Delta}}\frac{z^{2\Delta-d}}{(z^{2}+\left|x-x^{\prime}\right|^{2})^{\Delta}}.\label{eq: delta_function}
\end{equation}
By the mapping (\ref{eq:field_map}), the boundary values $u_{\pm}$ of the eigenmodes are connected to the boundary values $\phi_{1(0)}$ and $\phi_{2(0)}$ of the original fields by 
\begin{equation}
u_{\pm}=\frac{1}{2}\left(\phi_{1(0)}\pm\sqrt{\epsilon}\phi_{2(0)}\right).\label{eq:sources_pm}
\end{equation}
Here and below, the $\phi_{a(0)}$ defined by (\ref{eq:sources_pm}) play the role of the sources.

In terms of the original fields, we construct the bulk-to-boundary propagator as 
\begin{align}
\phi_{1}(z,x^{i}) & =\int d^{d}x^{\prime}\left(K_{11}(z,x^{i};0,x^{i\prime})\phi_{1(0)}(x^{i\prime})+K_{21}(z,x^{i};0,x^{i\prime})\phi_{2(0)}(x^{i\prime})\right),\\
\phi_{2}(z,x^{i}) & =\int d^{d}x^{\prime}\left(K_{12}(z,x^{i};0,x^{i\prime})\phi_{1(0)}(x^{i\prime})+K_{22}(z,x^{i};0,x^{i\prime})\phi_{2(0)}(x^{i\prime})\right).
\end{align}
From (\ref{eq:field_map}), (\ref{eq:propagator}) and (\ref{eq:sources_pm}), we find 
\begin{equation}
K_{11}=K_{22}=\frac{1}{2}\left(K_{\Delta_{+}}+K_{\Delta_{-}}\right),
\quad
\epsilon^{-1/2}K_{21}=\epsilon^{1/2}K_{12}=\frac{1}{2}\left(K_{\Delta_{+}}-K_{\Delta_{-}}\right),\label{eq:Kij}
\end{equation}
with boundary limit given by 
\begin{align}
\lim_{z\rightarrow0}\phi_{1}(z,x^{i})= & \lim_{z\rightarrow0}\left(\frac{z^{d-\Delta_{-}}+z^{d-\Delta_{+}}}{2}\phi_{1(0)}(x^{i})+\sqrt{\epsilon}\frac{-z^{d-\Delta_{-}}+z^{d-\Delta_{+}}}{2}\phi_{2(0)}(x^{i})\right),\\
\lim_{z\rightarrow0}\phi_{2}(z,x^{i})= & \lim_{z\rightarrow0}\left(\frac{1}{\sqrt{\epsilon}}\frac{-z^{d-\Delta_{-}}+z^{d-\Delta_{+}}}{2}\phi_{1(0)}(x^{i})+\frac{z^{d-\Delta_{-}}+z^{d-\Delta_{+}}}{2}\phi_{2(0)}(x^{i})\right).
\end{align}

By integrating by parts and substituting the equations of motion into the action, we can find the on-shell action.
Because of potential divergences, the action is regularized in the vicinity of the boundary \cite{GHEZELBASH1999IJMPA,KOGAN1999PLB,Bergshoeff2012JHEP}, 
\begin{equation}
S=-\frac{1}{2}\lim_{\zeta\rightarrow0}\int_{z=\zeta}d^{d}x\sqrt{\gamma}\left(\phi_{1}(\overrightarrow{n}\cdot\overrightarrow{\nabla})\phi_{2}+\phi_{2}(\overrightarrow{n}\cdot\overrightarrow{\nabla})\phi_{1}\right),\label{eq:boundary_action}
\end{equation}
where $\sqrt{\gamma}=z^{-d}$, $\overrightarrow{n}\cdot\overrightarrow{\nabla}=-z\partial_{z}$.
In terms of the eigenmodes, $\phi_{1}\phi_{2}=\epsilon^{-1/2}(\psi_{+}^{2}-\psi_{-}^{2})$.
The on-shell action is therefore the difference of the on-shell actions of two decoupled scalar fields, in which all terms mixing $\Delta_{+}$ and $\Delta_{-}$ cancel.
Removing the divergent contact terms by the counterterm generalized from ref.~\cite{Bergshoeff2011JHEP},
\begin{equation}
S_{\text{ct}}=\int_{z=\zeta}d^{d}x\sqrt{\gamma}\left(\frac{\phi_{1}^{2}+\epsilon\phi_{2}^{2}}{2\beta}-(d-\bar{\Delta})\phi_{1}\phi_{2}\right),\label{eq:counterterm}
\end{equation}
we obtain (see appendix~\ref{app: epsilon} for details) 
\begin{equation}
S=\frac{2}{\sqrt{\epsilon}}\int d^{d}xd^{d}x^{\prime}\left(\tilde{A}_{+}\frac{u_{+}(x)u_{+}(x^{\prime})}{\left|x-x^{\prime}\right|^{2\Delta_{+}}}-\tilde{A}_{-}\frac{u_{-}(x)u_{-}(x^{\prime})}{\left|x-x^{\prime}\right|^{2\Delta_{-}}}\right),\label{eq:PT_regularizedPaction}
\end{equation}
where $A_{\pm}\equiv\frac{2\Delta_{\pm}-d}{2}$ and $\tilde{A}_{\pm}\equiv A_{\pm}/N_{\Delta_{\pm}}$.
Each term here corresponds to a renormalized on-shell action of a decoupled field $\psi_{\pm}$ \cite{Kostas2002CQG}.
The two terms have opposite signs, as the two modes in (\ref{eq:action_in_psi}).
The counterterm used in holographic renormalization \cite{Kostas2002CQG} removes the divergent contact terms.
Its finite part fixes the normalization of the two-point functions at separated points.

\subsection{Two-point functions}

Following the AdS/CFT correspondence \cite{GUBSER1998PLB,witten1998ATMP}, the classical sources $\phi_{a(0)}$ couple to quantum local CFT operators $\mathscr{O}_{a}$ as
\begin{equation}
\left\langle \exp\int d^{d}x(\phi_{1(0)}\mathscr{O}_{1}+\beta\phi_{2(0)}\mathscr{O}_{2})\right\rangle _{\text{CFT}} = \left.e^{-S[\phi_{1},\phi_{2}]}\right|_{\phi_{1}(\partial\text{AdS})=\phi_{1(0)},\phi_{2}(\partial\text{AdS})=\phi_{2(0)}},\label{eq:GKPW}
\end{equation}
where $S$ is the renormalized on-shell action in (\ref{eq:PT_regularizedPaction}). 
The constant $\beta$ fixes the normalization of $\mathscr{O}_{2}$ relative to $\mathscr{O}_{1}$.
We drop hats on these local quantum operators and conformal generators for simplicity. 
We take $\beta=\Delta_{+}+\Delta_{-}-d$ as in (\ref{eq:eta_epsilon}), so that the dilatation operator has a unit off-diagonal entry (see section~\ref{sec:CFT_calculation}).
Equivalently, by (\ref{eq:sources_pm}), $\mathscr{O}_{1}=(\mathscr{O}_{+}+\mathscr{O}_{-})/2$ and $\beta\mathscr{O}_{2}=\sqrt{\epsilon}(\mathscr{O}_{+}-\mathscr{O}_{-})/2$, where $\mathscr{O}_{\pm}$ are the operators dual to the eigenmodes $\psi_{\pm}$.
The resulting two-point functions are, with $\sepdist\equiv\left|x_{1}-x_{2}\right|$ (see appendix~\ref{app: epsilon}), 
\begin{equation}
\begin{aligned}
\langle\mathscr{O}_{1}(x_{1})\mathscr{O}_{1}(x_{2})\rangle & =\frac{1}{\sqrt{\epsilon}}\left(\frac{\tilde{A}_{-}}{\sepdist^{2\Delta_{-}}}-\frac{\tilde{A}_{+}}{\sepdist^{2\Delta_{+}}}\right),\\
\beta\langle\mathscr{O}_{1}(x_{1})\mathscr{O}_{2}(x_{2})\rangle & =-\left(\frac{\tilde{A}_{-}}{\sepdist^{2\Delta_{-}}}+\frac{\tilde{A}_{+}}{\sepdist^{2\Delta_{+}}}\right),\\
\beta^{2}\langle\mathscr{O}_{2}(x_{1})\mathscr{O}_{2}(x_{2})\rangle & =\sqrt{\epsilon}\left(\frac{\tilde{A}_{-}}{\sepdist^{2\Delta_{-}}}-\frac{\tilde{A}_{+}}{\sepdist^{2\Delta_{+}}}\right).
\end{aligned}
\label{eq:AdS_2pt_powerlaws}
\end{equation}
Writing $\sepdist^{-2\Delta_{\pm}}=\sepdist^{-2\bar{\Delta}}e^{\mp2\delta\ln \sepdist}$ and using $\sqrt{\epsilon}=\beta\delta$ from (\ref{eq:eta_epsilon}), they become 
\begin{align}
\langle\mathscr{O}_{1}(x_{1})\mathscr{O}_{1}(x_{2})\rangle & =\frac{1}{\sepdist^{2\bar{\Delta}}}\left[c_{1}\cosh(2\delta\ln \sepdist)-c_{2}\delta^{-1}\sinh(2\delta\ln \sepdist)\right],\label{eq:PT_AdS_2pt_1}\\
\langle\mathscr{O}_{1}(x_{1})\mathscr{O}_{2}(x_{2})\rangle & =\frac{1}{\sepdist^{2\bar{\Delta}}}\left[c_{2}\cosh(2\delta\ln \sepdist)-c_{1}\delta\sinh(2\delta\ln \sepdist)\right],\label{eq:PT_AdS_2pt_2}\\
\langle\mathscr{O}_{2}(x_{1})\mathscr{O}_{2}(x_{2})\rangle & =\eta\langle\mathscr{O}_{1}(x_{1})\mathscr{O}_{1}(x_{2})\rangle=\frac{1}{\sepdist^{2\bar{\Delta}}}\left[\eta c_{1}\cosh(2\delta\ln \sepdist)-c_{2}\delta\sinh(2\delta\ln \sepdist)\right],\label{eq:PT_AdS_2pt_3}
\end{align}
with 
\begin{equation}
c_{1}\equiv-\frac{\tilde{A}_{+}-\tilde{A}_{-}}{\sqrt{\epsilon}},\quad c_{2}\equiv-\frac{\tilde{A}_{+}+\tilde{A}_{-}}{\beta}.\label{eq:c1c2}
\end{equation}

\begin{figure}[t]
\centering
\includegraphics{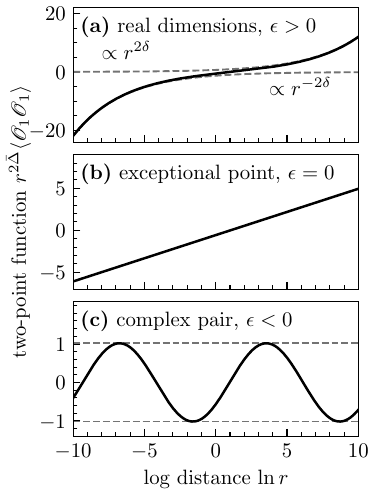}
\caption{Two-point function $\langle\mathscr{O}_{1}(x_{1})\mathscr{O}_{1}(x_{2})\rangle$, multiplied by $\sepdist^{2\bar{\Delta}}$, as a function of $\ln\sepdist$ with $\sepdist$ in units of the AdS radius, for $d=3$ and $m^{2}=-1$. \textbf{(a)} Real scaling dimensions, $\epsilon=0.1$: linear combination of two power laws. The dashed curves are the contributions proportional to $\sepdist^{-2\Delta_{+}}$ and $\sepdist^{-2\Delta_{-}}$. Multiplied by $\sepdist^{2\bar{\Delta}}$, they scale as $\sepdist^{-2\delta}$ and $\sepdist^{2\delta}$. \textbf{(b)} Exceptional point, $\epsilon=0$: linear in $\ln\sepdist$. \textbf{(c)} Complex-conjugate scaling dimensions, $\epsilon=-0.5$: log-periodic oscillation. The dashed lines are its envelope. In PT-symmetric theories with the same structure, (a) and (c) correspond to the PT-unbroken and PT-broken phases (section~\ref{sec: PT_transition}). See section~\ref{sec:AdS_calculation} for details.}
\label{fig:two_point}
\end{figure}

The two-point functions (\ref{eq:PT_AdS_2pt_1})--(\ref{eq:PT_AdS_2pt_3}) are real for both signs of $\epsilon$, with three distinct behaviors:
\begin{enumerate}
\item[(i)] $\epsilon>0$: the scaling dimensions are real, and the two-point functions are linear combinations of two power laws, $\sepdist^{-2\Delta_{+}}$ and $\sepdist^{-2\Delta_{-}}$.
\item[(ii)] $\epsilon=0$: the two scaling dimensions coalesce, and the two-point functions continuously reduce to those of the dipole model dual to an LCFT, as derived in the next subsection.
\item[(iii)] $\epsilon<0$: the scaling dimensions are complex conjugate, and the two-point functions exhibit periodic oscillations in $\ln\sepdist$.
\end{enumerate}
The three behaviors are illustrated in figure~\ref{fig:two_point}.
In section~\ref{sec:CFT_calculation}, we show that $\epsilon>0$ and $\epsilon<0$ correspond to real and complex-conjugate spectra of the dilatation operator, which plays the role of the Hamiltonian in radial quantization.
In section~\ref{sec: PT_transition}, we show that, in PT-symmetric theories with the same structure, they are the PT-unbroken and PT-broken phases.

\subsection{Approaching \texorpdfstring{$\epsilon=0$}{epsilon = 0}}

Notice that, as $\epsilon\rightarrow0$, the prefactor $\epsilon^{-1/2}$ in (\ref{eq:AdS_2pt_powerlaws}) diverges.
This raises the question of whether the $\epsilon\rightarrow0$ limit is finite, and how it is related to the dipole model of (\ref{eq:dipole_model}). 

We work in the form (\ref{eq:PT_AdS_2pt_1})--(\ref{eq:PT_AdS_2pt_3}). 
As $\epsilon\rightarrow0$, $\delta\rightarrow0$, so that $\cosh(2\delta\ln \sepdist)\rightarrow1$, $\sinh(2\delta\ln \sepdist)/\delta\rightarrow2\ln \sepdist$ and $\delta\sinh(2\delta\ln \sepdist)\rightarrow0$.
In addition, $\bar{\Delta}\rightarrow\Delta$ and $\beta\rightarrow\sqrt{d^{2}+4m^{2}}$.
The function $\tilde{A}=\frac{2\Delta-d}{2N_{\Delta}}$ is differentiable in $\Delta$ for $\Delta>d/2$, since the gamma function is.
With $\Delta_{\pm}=\bar{\Delta}\pm\delta$ and $\delta=\sqrt{\epsilon}/\beta$, we can write 
\begin{equation}
\frac{\tilde{A}_{+}-\tilde{A}_{-}}{\sqrt{\epsilon}}\stackrel{\epsilon\rightarrow0}{=}\partial_{\Delta}\tilde{A}\frac{\Delta_{+}-\Delta_{-}}{\sqrt{\epsilon}}\stackrel{\epsilon\rightarrow0}{=}\frac{2\partial_{\Delta}\tilde{A}}{\sqrt{d^{2}+4m^{2}}},
\qquad
\tilde{A}_{+}+\tilde{A}_{-}\stackrel{\epsilon\rightarrow0}{=}2\tilde{A}.
\end{equation}
Hence, we have $c_{1}\rightarrow-2\partial_{\Delta}\tilde{A}/\sqrt{d^{2}+4m^{2}}$ and $c_{2}\rightarrow-2\tilde{A}/\sqrt{d^{2}+4m^{2}}$.
The resulting non-vanishing two-point functions are 
\begin{align}
\langle\mathscr{O}_{1}(x_{1})\mathscr{O}_{1}(x_{2})\rangle= & \frac{2\tilde{A}}{\sqrt{d^{2}+4m^{2}}}\frac{2\ln\sepdist}{\sepdist^{2\Delta}}-\frac{2\partial_{\Delta}\tilde{A}}{\sqrt{d^{2}+4m^{2}}}\frac{1}{\sepdist^{2\Delta}},\label{eq:finite_2ptFunction_1}\\
\beta\langle\mathscr{O}_{1}(x_{1})\mathscr{O}_{2}(x_{2})\rangle= & -\frac{2\tilde{A}}{\sepdist^{2\Delta}},\label{eq:finite_2ptFunction_2}
\end{align}
and $\langle\mathscr{O}_{2}(x_{1})\mathscr{O}_{2}(x_{2})\rangle=0$.
Notably, these two-point functions coincide with those of the dipole model obtained by a direct calculation at $\epsilon=0$ in appendix~\ref{app:dipole}, (\ref{eq:diverging_2ptFunction_1})--(\ref{eq:diverging_2ptFunction_2}).
Therefore, the two-point functions are continuous around $\epsilon=0$, where the two power laws merge into the logarithmic term of $\ln\sepdist$.
The logarithmic term identifies the dual CFT as an LCFT.
In section~\ref{sec:CFT_calculation}, we show that these two-point functions match those of an LCFT.

\section{Two-point functions on the CFT side}\label{sec:CFT_calculation}

In this section, we show that conformal symmetry alone fixes the form of the holographic two-point functions of section~\ref{sec:AdS_calculation}, for any pair of scalar primary operators on which the dilatation operator acts as a real $2\times2$ matrix.
We start from the non-diagonalizable dilatation operator of an LCFT and deform it by an $\eta$ term, which controls the real-to-complex transition in its spectrum.
We then derive the two-point functions for all $\eta$ from the conformal Ward identities and match them with the holographic results (\ref{eq:PT_AdS_2pt_1})--(\ref{eq:PT_AdS_2pt_3}).
This matching identifies the dilatation operator, i.e., the Hamiltonian in radial quantization, of the dual CFT.

\subsection{Dilatation operator of an LCFT and its deformation}

Recall that the quantum conformal generators act on the operator multiplet $\Phi(x)=(\mathscr{O}_{1}(x),\mathscr{O}_{2}(x))^{T}$ by commutators \cite{francesco2012conformal},
\begin{align}
[P_{\mu},\Phi(x)] & =-i\partial_{\mu}\Phi(x),\\
[L_{\mu\nu},\Phi(x)] & =i(x_{\mu}\partial_{\nu}-x_{\nu}\partial_{\mu})\Phi(x)+\tilde{L}_{\mu\nu}\Phi(x),\\
[D,\Phi(x)] & =(-ix^{\nu}\partial_{\nu}+\tilde{D})\Phi(x),\\
[K_{\mu},\Phi(x)] & =(\tilde{K}_{\mu}+2x_{\mu}\tilde{D}-2x^{\nu}\tilde{L}_{\mu\nu}-2ix_{\mu}x^{\nu}\partial_{\nu}+ix^{2}\partial_{\mu})\Phi(x),
\end{align}
where $\tilde{L}_{\mu\nu}$, $\tilde{D}$ and $\tilde{K}_{\mu}$ are the matrix representations of $L_{\mu\nu}$, $D$ and $K_{\mu}$ at $x=0$.
Assuming $\Phi(x)$ belongs to an \emph{irreducible} representation of the Lorentz group, we find that the scaling dimension $\Delta$ is given by $i\tilde{D}=\Delta I$ with the identity matrix $I$, as $\tilde{D}$ commutes with all the generators of the Lorentz group.

To identify the quantum Hamiltonian, write the Euclidean radial coordinate as $\varrho=|x|=R e^{\tau/R}$.
The flat metric becomes
\begin{equation}
ds_{\mathbb{R}^{d}}^{2}=d\varrho^{2}+\varrho^{2}d\Omega_{d-1}^{2}
=e^{2\tau/R}(d\tau^{2}+R^{2}d\Omega_{d-1}^{2}).
\end{equation}
Multiplying the metric by the position-dependent factor $e^{-2\tau/R}$ removes the overall prefactor and gives the cylinder metric $ds_{\text{cyl}}^{2}=d\tau^{2}+R^{2}d\Omega_{d-1}^{2}$ on $\mathbb{R}_{\tau}\times S^{d-1}_{R}$. 
A dilation $\varrho\mapsto e^{\alpha}\varrho$ becomes a Euclidean-time translation $\tau\mapsto\tau+R\alpha$.
Thus, in our convention, the quantum Hamiltonian is
\begin{equation}
\hat{H}=\frac{iD}{R},\qquad e^{-\tau\hat{H}}=e^{-(\tau/R)iD}.
\label{eq:H_dilatation}
\end{equation}
Here, $D$ is the quantum generator, with its hat suppressed.
After Wick rotation $\tau=it$, $\hat{H}$ generates the time evolution $e^{-it\hat{H}}$ on the spatial sphere.
The state--operator correspondence gives $|\mathscr{O}_{a}\rangle=\mathscr{O}_{a}(0)|0\rangle$.
With $D|0\rangle=0$, the action on these primary states is
\begin{equation}
\hat{H}|\mathscr{O}_{a}\rangle=\frac{iD}{R}|\mathscr{O}_{a}\rangle
=\frac{1}{R}[iD,\mathscr{O}_{a}(0)]|0\rangle
=\frac{1}{R}(i\tilde{D})_{a}{}^{b}|\mathscr{O}_{b}\rangle.
\label{eq:H_primary_sector}
\end{equation}
Thus, $i\tilde{D}$ represents the quantum Hamiltonian in the primary sector.
A scaling dimension $\Delta$ gives an excitation energy $E=\Delta/R$.
We use $R=1$ and $\hbar=c=1$ throughout. 

An LCFT is characterized by the same conformal group, but $\Phi(x)$ belongs to a \emph{reducible} representation of the Lorentz group.
Schur's lemma \cite{francesco2012conformal} then does not force $i\tilde{D}$ to be proportional to the identity.
In a non-unitary theory, $i\tilde{D}$ need not be diagonalizable.
The multiplet $\Phi(x)$ then forms a \emph{reducible but indecomposable} representation of the conformal group \cite{Hogervorst2017JHEP}.
Choosing two states $\{|\mathscr{O}_{1}\rangle,|\mathscr{O}_{2}\rangle\}$ as the basis, we can write $\tilde{D}$ in Jordan normal form, 
\begin{equation}
iD|\mathscr{O}_{a}\rangle=(i\tilde{D})_{a}{}^{b}|\mathscr{O}_{b}\rangle,\,\,\,i\tilde{D}=\left(\begin{array}{cc}
\Delta & 1\\
0 & \Delta
\end{array}\right).\label{eq: non-diagonalizable_dilitation}
\end{equation}
In this basis, $\mathscr{O}_{2}$ is the eigenoperator, $iD|\mathscr{O}_{2}\rangle=\Delta|\mathscr{O}_{2}\rangle$, and $\mathscr{O}_{1}$ is its logarithmic partner.
We now deform this Jordan block by an $\eta$ term, which splits the two coalescing scaling dimensions.
In the same basis $\{|\mathscr{O}_{1}\rangle,|\mathscr{O}_{2}\rangle\}$, the modified form of $i\tilde{D}(\eta)$ is given by 
\begin{equation}
i\tilde{D}(\eta)=\left(\begin{array}{cc}
\bar{\Delta} & 1\\
\eta & \bar{\Delta}
\end{array}\right),\label{eq:perturbed_D}
\end{equation}
where $\bar{\Delta}$ is the mean of the two eigenvalues, $\bar{\Delta}\pm\sqrt{\eta}$, and $\bar{\Delta}=\Delta$ at $\eta=0$.
This specific form ensures that the eigenvalues exhibit a real-to-complex transition as we tune $\eta$ from positive to negative values, with the exceptional point at $\eta=0$, where (\ref{eq:perturbed_D}) reduces to (\ref{eq: non-diagonalizable_dilitation}).
This form is general: every real dilatation matrix of a pair of scalar primary operators that is not proportional to the identity takes the form (\ref{eq:perturbed_D}) in a suitable real basis (see appendix~\ref{app:generality}).
The commutation relations of conformal generators give rise to the following reduced algebra at the origin: 
\begin{align}
 & [\tilde{D},\tilde{K}_{\mu}]=-i\tilde{K}_{\mu},\label{eq: reduced_commulation_relation_1}\\{}
[\tilde{D},\tilde{L}_{\mu\nu}]=0,\,\,\, & [\tilde{L}_{\mu\nu},\tilde{L}_{\rho\sigma}]=i(\delta_{\rho\nu}\tilde{L}_{\mu\sigma}+\delta_{\mu\sigma}\tilde{L}_{\nu\rho}-\delta_{\mu\rho}\tilde{L}_{\nu\sigma}-\delta_{\nu\sigma}\tilde{L}_{\mu\rho}).\label{eq: reduced_commulation_relation_2}
\end{align}
Since we consider scalar operators, $\tilde{L}_{\mu\nu}=0$.
The commutator $[\tilde{K}_{\mu},\tilde{L}_{\nu\rho}]\propto\delta_{\mu\nu}\tilde{K}_{\rho}-\delta_{\mu\rho}\tilde{K}_{\nu}$ then gives $\tilde{K}_{\mu}=0$ for $d\geq2$, for every $\eta$.

\subsection{Ward identities and two-point functions}

Now, we proceed to determine the quantum two-point functions for scalar operators $\mathscr{O}_{a}(x)$ by assuming that conformal symmetry is not broken at the quantum level.
Beginning with translations, the Ward identity for translations states that
\begin{equation}
\left(\frac{\partial}{\partial x_{1}^{\mu}}+\frac{\partial}{\partial x_{2}^{\mu}}\right)\langle\mathscr{O}_{a}(x_{1})\mathscr{O}_{b}(x_{2})\rangle=0.
\end{equation}
This implies that 
\begin{equation}
\langle\mathscr{O}_{a}(x_{1})\mathscr{O}_{b}(x_{2})\rangle=f(x_{1}-x_{2}),
\end{equation}
where $f(x)$ is a generic function of $x$.
A finite Lorentz transformation $x_{\mu}^{\prime}=\Lambda_{\mu}^{\nu}x_{\nu}$ means that
\begin{equation}
\langle\mathscr{O}_{a}(x_{1}^{\mu})\mathscr{O}_{b}(x_{2}^{\mu})\rangle=\langle\mathscr{O}_{a}(\Lambda_{\nu}^{\mu}x_{1}^{\nu})\mathscr{O}_{b}(\Lambda_{\nu}^{\mu}x_{2}^{\nu})\rangle.
\end{equation}
This leads to 
\begin{equation}
\langle\mathscr{O}_{a}(x_{1})\mathscr{O}_{b}(x_{2})\rangle=f(|x_{1}-x_{2}|).\label{eq: 2ptFunction_translation_Lorentz}
\end{equation}
Similarly, a scaling transformation acts by conjugation with $U_{\alpha}=e^{i\alpha D}$,
\begin{equation}
U_{\alpha}\mathscr{O}_{a}(x)U_{\alpha}^{-1}
=\left[e^{\alpha i\tilde{D}}\right]_{a}^{b}\mathscr{O}_{b}(e^{\alpha}x).
\end{equation}
Vacuum invariance then gives
\begin{align}
\langle\mathscr{O}_{a}(x_{1})\mathscr{O}_{b}(x_{2})\rangle & =\langle (U_{\alpha}\mathscr{O}_{a}(x_{1})U_{\alpha}^{-1})(U_{\alpha}\mathscr{O}_{b}(x_{2})U_{\alpha}^{-1})\rangle\nonumber \\
 & =\langle\left[e^{\alpha i\tilde{D}}\right]_{a}^{a^{\prime}}\mathscr{O}_{a^{\prime}}(e^{\alpha}x_{1})\left[e^{\alpha i\tilde{D}}\right]_{b}^{b^{\prime}}\mathscr{O}_{b^{\prime}}(e^{\alpha}x_{2})\rangle.\label{eq: 2pt_scaling_transformation}
\end{align}
We see that if $i\tilde{D}=\Delta I$, the above relation leads to $f(|x_{1}-x_{2}|)=(e^{\alpha})^{2\Delta}f(e^{\alpha}|x_{1}-x_{2}|)$.
We thus obtain the standard two-point functions $\langle\mathscr{O}_{a}(x_{1})\mathscr{O}_{b}(x_{2})\rangle=\frac{c}{|x_{1}-x_{2}|^{2\Delta}}$.

For a general matrix $i\tilde{D}$, we write the two-point functions (\ref{eq: 2ptFunction_translation_Lorentz}) as a matrix, $G_{ab}(\sepdist)\equiv\langle\mathscr{O}_{a}(x_{1})\mathscr{O}_{b}(x_{2})\rangle$ with $\sepdist\equiv|x_{1}-x_{2}|$, which is symmetric, $G_{ab}=G_{ba}$, for bosonic operators.
Differentiating (\ref{eq: 2pt_scaling_transformation}) with respect to $\alpha$ at $\alpha=0$ gives 
\begin{equation}
\sepdist\frac{dG}{d\sepdist}=-i\tilde{D}G-G(i\tilde{D})^{T}.\label{eq: dilatation_Ward}
\end{equation}
For special conformal transformations, we set $x_{1}=0$, where $[K_{\mu},\mathscr{O}_{a}(0)]=0$ since $\tilde{K}_{\mu}=0$.
With $\tilde{L}_{\mu\nu}=0$, the Ward identity $\langle\mathscr{O}_{a}(0)[K_{\mu},\mathscr{O}_{b}(x_{2})]\rangle=0$ then gives $\sepdist\frac{dG}{d\sepdist}=-2G(i\tilde{D})^{T}$, with $\tilde{D}$ acting on the second index.
The difference between this identity and (\ref{eq: dilatation_Ward}) gives 
\begin{equation}
i\tilde{D}G=G(i\tilde{D})^{T}.\label{eq:SCT_constraint}
\end{equation}
Combining (\ref{eq: dilatation_Ward}) and (\ref{eq:SCT_constraint}), we obtain $\sepdist\,dG/d\sepdist=-2i\tilde{D}G$, whose solution is 
\begin{equation}
G(\sepdist)=\sepdist^{-2i\tilde{D}}G(1),\quad\sepdist^{-2i\tilde{D}}\equiv e^{-2i\tilde{D}\ln\sepdist}.\label{eq: G_solution}
\end{equation}
For (\ref{eq:perturbed_D}), the $(1,2)$ component of (\ref{eq:SCT_constraint}), $\bar{\Delta}G_{12}+G_{22}=\eta G_{11}+\bar{\Delta}G_{12}$, gives $G_{22}=\eta G_{11}$, so that 
\begin{equation}
G(1)=\left(\begin{array}{cc}
c_{1} & c_{2}\\
c_{2} & \eta c_{1}
\end{array}\right),\label{eq:G1}
\end{equation}
with two normalization constants $c_{1}$ and $c_{2}$.
Writing $i\tilde{D}(\eta)=\bar{\Delta}I+N$ with $N\equiv\left(\begin{array}{cc}
0 & 1\\
\eta & 0
\end{array}\right)$ and $N^{2}=\eta I$, we have 
\begin{equation}
\sepdist^{-2i\tilde{D}(\eta)}=\sepdist^{-2\bar{\Delta}}\left(\cosh(2\sqrt{\eta}\ln\sepdist)\,I-\frac{\sinh(2\sqrt{\eta}\ln\sepdist)}{\sqrt{\eta}}N\right).
\end{equation}
The complete forms of two-point functions are thus given by 
\begin{align}
\langle\mathscr{O}_{1}(x_{1})\mathscr{O}_{1}(x_{2})\rangle & =\frac{1}{\sepdist^{2\bar{\Delta}}}\left[c_{1}\cosh(2\sqrt{\eta}\ln\sepdist)-c_{2}\frac{1}{\sqrt{\eta}}\sinh(2\sqrt{\eta}\ln\sepdist)\right],\label{eq:PT_EP_2pt_1}\\
\langle\mathscr{O}_{1}(x_{1})\mathscr{O}_{2}(x_{2})\rangle & =\frac{1}{\sepdist^{2\bar{\Delta}}}\left[c_{2}\cosh(2\sqrt{\eta}\ln\sepdist)-c_{1}\sqrt{\eta}\sinh(2\sqrt{\eta}\ln\sepdist)\right],\label{eq:PT_EP_2pt_2}\\
\langle\mathscr{O}_{2}(x_{1})\mathscr{O}_{2}(x_{2})\rangle & =\frac{1}{\sepdist^{2\bar{\Delta}}}\left[\eta c_{1}\cosh(2\sqrt{\eta}\ln\sepdist)-c_{2}\sqrt{\eta}\sinh(2\sqrt{\eta}\ln\sepdist)\right].\label{eq:PT_EP_2pt_3}
\end{align}
For $\eta>0$, they are linear combinations of two power laws, $\sepdist^{-2(\bar{\Delta}\pm\sqrt{\eta})}$.
For $\eta<0$, $\cosh(2\sqrt{\eta}\ln\sepdist)=\cos(2\sqrt{|\eta|}\ln\sepdist)$ and $\sinh(2\sqrt{\eta}\ln\sepdist)/\sqrt{\eta}=\sin(2\sqrt{|\eta|}\ln\sepdist)/\sqrt{|\eta|}$, so that they are real and exhibit log-periodic oscillations.
At $\eta=0$, $\cosh(2\sqrt{\eta}\ln\sepdist)\rightarrow1$ and $\sinh(2\sqrt{\eta}\ln\sepdist)/\sqrt{\eta}\rightarrow2\ln\sepdist$.
We then find the two-point functions of the LCFT (see also \cite{GURARIE1993NPB,CAUX1996NPB}) 
\begin{align}
\langle\mathscr{O}_{1}(x_{1})\mathscr{O}_{1}(x_{2})\rangle & =\frac{-2c_{2}\ln|x_{1}-x_{2}|}{|x_{1}-x_{2}|^{2\Delta}}+\frac{c_{1}}{|x_{1}-x_{2}|^{2\Delta}},\label{eq: EP_2pt_1}\\
\langle\mathscr{O}_{1}(x_{1})\mathscr{O}_{2}(x_{2})\rangle & =\frac{c_{2}}{|x_{1}-x_{2}|^{2\Delta}}=\langle\mathscr{O}_{2}(x_{1})\mathscr{O}_{1}(x_{2})\rangle,\label{eq: EP_2pt_2}\\
\langle\mathscr{O}_{2}(x_{1})\mathscr{O}_{2}(x_{2})\rangle & =0.\label{eq: EP_2pt_3}
\end{align}
The logarithmic term gives LCFTs their name.

The solution (\ref{eq: G_solution}) also provides a direct way to extract the spectrum of the dilatation operator from the two-point functions: for two nearby separations $\sepdist\neq\sepdist_{0}$ and $\det G(1)\neq0$, $G(\sepdist)G(\sepdist_{0})^{-1}=(\sepdist/\sepdist_{0})^{-2i\tilde{D}}$, whose eigenvalues are $(\sepdist/\sepdist_{0})^{-2\Delta_{\pm}}$ with $\Delta_{\pm}=\bar{\Delta}\pm\sqrt{\eta}$.
The eigenvalues are real for $\eta>0$ and complex conjugate for $\eta<0$.
At $\eta=0$, the matrix $G(\sepdist)G(\sepdist_{0})^{-1}$ is itself a non-diagonalizable Jordan block.
Thus, the phase and the exceptional point can be identified from a $2\times2$ matrix of two-point functions at two separations, without fitting their individual components.

\subsection{Matching the AdS side}

The holographic two-point functions (\ref{eq:PT_AdS_2pt_1})--(\ref{eq:PT_AdS_2pt_3}) take exactly the form (\ref{eq:PT_EP_2pt_1})--(\ref{eq:PT_EP_2pt_3}), with $\sqrt{\eta}=\delta$.
The operator basis fixed by (\ref{eq:GKPW}) and (\ref{eq:sources_pm}) gives the dilatation operator in the form (\ref{eq:perturbed_D}), with $\eta=\epsilon/\beta^{2}$ as in (\ref{eq:eta_epsilon}).
The only remaining identifications are the constants, 
\begin{equation}
c_{1}\leftrightarrow-\frac{\tilde{A}_{+}-\tilde{A}_{-}}{\sqrt{\epsilon}},\quad c_{2}\leftrightarrow-\frac{\tilde{A}_{+}+\tilde{A}_{-}}{\beta},\label{eq: c_identification}
\end{equation}
which are (\ref{eq:c1c2}).
At $\epsilon=0$, they reduce to $c_{1}\leftrightarrow-2\partial_{\Delta}\tilde{A}/\beta$ and $c_{2}\leftrightarrow-2\tilde{A}/\beta$ with $\beta=\sqrt{d^{2}+4m^{2}}$.
The holographic two-point functions of the dipole model (\ref{eq:diverging_2ptFunction_1})--(\ref{eq:diverging_2ptFunction_2}), which coincide with the $\epsilon\rightarrow0$ limit (\ref{eq:finite_2ptFunction_1})--(\ref{eq:finite_2ptFunction_2}), exactly match (\ref{eq: EP_2pt_1})--(\ref{eq: EP_2pt_3}) \cite{GHEZELBASH1999IJMPA,Bergshoeff2011JHEP}.
Therefore, the matrix $i\tilde{D}$ of the dual quantum CFT takes the form (\ref{eq:perturbed_D}) in the two-primary sector.
By (\ref{eq:H_primary_sector}), its eigenvalues are the excitation energies of $\hat{H}$.
They are real for $\epsilon>0$ ($\eta>0$) and complex conjugate for $\epsilon<0$ ($\eta<0$).
In these two spectral phases, the quantum two-point functions are linear combinations of two power laws and exhibit log-periodic oscillations, respectively.

\section{PT symmetry-breaking transition}\label{sec: PT_transition}

This section explains the relation between the real-to-complex transition of the scaling dimensions and PT symmetry breaking.
Under the assumptions stated below, time reversal provides an antilinear quantum symmetry that the energy eigenstates preserve or break. 
We then discuss the real-time signatures of the spectral transition, including the thermal response.

\subsection{From real-to-complex transition to PT transition}

By (\ref{eq:H_dilatation}) and (\ref{eq:H_primary_sector}), $i\tilde{D}(\eta)$ represents the quantum Hamiltonian in the two-primary sector, with excitation energies $E_{\pm}=\Delta_{\pm}=\bar{\Delta}\pm\sqrt{\eta}$.
At $\epsilon=\eta=0$, this sector is a non-diagonalizable Jordan block, the exceptional point \cite{Ashida2020AP} characteristic of an LCFT \cite{Hogervorst2017JHEP}.
The deformation produces a real-to-complex spectral transition \cite{Wang2023arXivCLEP}.
The inner products of the two primary states are given by $G(1)$ in (\ref{eq:G1}), whose determinant $\eta c_{1}^{2}-c_{2}^{2}=-4\tilde{A}_{+}\tilde{A}_{-}/\beta^{2}$ is negative for every $\epsilon$.
One state therefore has negative norm, like the bulk mode noted in section~\ref{sec:model}.
The dual CFT is thus non-unitary, as is typical for LCFTs \cite{Hogervorst2017JHEP,Bergshoeff2012JHEP}.
Nevertheless, the constraint (\ref{eq:SCT_constraint}) implies that $\hat{H}$ is self-adjoint with respect to this indefinite inner product.
The Hamiltonian sector is therefore pseudo-Hermitian for every $\epsilon$ \cite{Mostafazadeh2002JMP}.
An antilinear symmetry such as PT also constrains the spectrum to real eigenvalues or complex-conjugate pairs \cite{bender2023arXiv}.
PT symmetry is unbroken when the energy eigenstates can be chosen PT invariant, and broken when PT exchanges eigenstates with distinct complex-conjugate energies \cite{Ashida2020AP}.
Identification of the PT transition requires verification of two criteria: (i) the antilinear symmetry is a physical symmetry of the theory, and (ii) the eigenstates in the real and complex spectral phases correspondingly preserve and break this symmetry.

For the first criterion, we consider the Lorentzian continuation of the action (\ref{eq:the_action}) on global AdS$_{d+1}$, $ds^{2}=\cos^{-2}\rho\,(-dt^{2}+d\rho^{2}+\sin^{2}\rho\,d\Omega_{d-1}^{2})$.
The classical action is invariant under $t\mapsto-t$ for every real $\epsilon$.
On complex mode solutions, time reversal acts as
\begin{equation}
(\mathcal{T}f)(t,\rho,\Omega)=f^{*}(-t,\rho,\Omega).
\label{eq:classical_time_reversal}
\end{equation}
We then observe that $f_a^{*}(-t,\rho,\Omega)$ satisfies the same equations (\ref{eq:PT_modified_EOM}) as $f_a(t,\rho,\Omega)$. 
Assuming that the boundary conditions and quantization preserve this symmetry, it has an antilinear quantum implementation $\hat{T}$.
We choose parity-even scalar operators, so that $\hat{P}$ acts trivially and $\hat{\Theta}\equiv\hat{P}\hat{T}$ reduces to time reversal, with $\hat{\Theta}\hat{H}\hat{\Theta}^{-1}=\hat{H}$.

We further assume that $\hat{\Theta}$ leaves the operators dual to the real fields $\phi_{a}$ invariant, $\hat{\Theta}\mathscr{O}_{a}(0)\hat{\Theta}^{-1}=\mathscr{O}_{a}(0)$.
Assuming also a PT-invariant vacuum, the state--operator correspondence then gives
\begin{equation}
\hat{\Theta}|\mathscr{O}_{a}\rangle
=\left(\hat{\Theta}\mathscr{O}_{a}(0)
\hat{\Theta}^{-1}\right)\hat{\Theta}|0\rangle
=|\mathscr{O}_{a}\rangle.
\end{equation}
Antilinearity gives 
\begin{equation}
\hat{\Theta}\left(a|\mathscr{O}_{1}\rangle
+b|\mathscr{O}_{2}\rangle\right)
=a^{*}|\mathscr{O}_{1}\rangle
+b^{*}|\mathscr{O}_{2}\rangle.
\end{equation}
With this construction, PT acts as complex conjugation in the two-primary basis.
The Hamiltonian restricted to this sector preserves PT.

For the second criterion, we examine the energy eigenstates.
Under the assumed quantum PT symmetry,
\begin{equation}
\hat{H}|E\rangle=E|E\rangle
\quad\Longrightarrow\quad
\hat{H}\hat{\Theta}|E\rangle=E^{*}\hat{\Theta}|E\rangle.
\end{equation}
The eigenstates of our two-primary sector are $|\mathscr{O}_{\pm}\rangle=|\mathscr{O}_{1}\rangle\pm|\mathscr{O}_{2}\rangle/\sqrt{\eta}$, created by the operators $\mathscr{O}_{\pm}$ dual to the eigenmodes $\psi_{\pm}$.
For $\eta>0$, their coefficients and energies are real, so that each state is PT invariant.
For $\eta<0$, their coefficients and energies are complex conjugate, so that $\hat{\Theta}$ exchanges the two states. 
In the bulk, $\hat{\Theta}$ correspondingly exchanges the eigenmodes $\psi_{\pm}=(\phi_{1}\pm\sqrt{\epsilon}\phi_{2})/2$, since $\sqrt{\epsilon}$ is imaginary.
At $\eta=0$, the eigenstates coalesce at the exceptional point.
Under these assumptions, $\epsilon>0$ and $\epsilon<0$ correspond to the PT-unbroken and PT-broken phases, respectively \cite{Ashida2020AP}.

\subsection{Connections to real-time dynamics}\label{sec:realtime}

The real-to-complex transition also has a counterpart in the classical real-time dynamics of the eigenmodes $\psi_{\pm}$.
In global AdS, the normalizable modes of the eigenmodes $\psi_{\pm}$, i.e., solutions without a source at the boundary, have frequencies \cite{Balasubramanian1999PRD} 
\begin{equation}
\omega=\Delta_{\pm}+\ell+2n,\qquad n=0,1,2,\dots,\label{eq: global_modes}
\end{equation}
with angular momentum $\ell$.
These classical frequencies match the excitation energies of the dual quantum CFT states \cite{Fitzpatrick2011JHEP}.
The modes with $n=\ell=0$ give the primary energies $E_{\pm}=\Delta_{\pm}$, and the remaining modes give the descendants.
In $d=2$, these energies correspond to the finite-size excitation gaps of a critical system, which are proportional to the scaling dimensions.
Complex scaling dimensions are identified in this way in a non-Hermitian quantum Potts chain \cite{Tang2024PRL}.
The map (\ref{eq:classical_time_reversal}) sends $e^{-i\omega t}u(\rho,\Omega)$ to $e^{-i\omega^{*}t}u^{*}(\rho,\Omega)$.
For $\epsilon>0$, all frequencies are real.
For $\epsilon<0$, the frequencies $\omega=\bar{\Delta}+\ell+2n\pm i\kappa$, with $\kappa\equiv\mathrm{Im}\,\Delta_{+}$, are complex conjugate.
The mode solution $e^{-i\omega t}$ with $\mathrm{Im}\,\omega>0$ grows as $e^{\kappa t}$, whereas the one with $\mathrm{Im}\,\omega<0$ decays as $e^{-\kappa t}$.
Under the symmetry assumptions stated above, this mode pairing reflects PT breaking of the dual quantum eigenstates, as in systems with balanced gain and loss \cite{Ashida2020AP}.
As $\epsilon\to0$, the two normalizable modes with fixed $\ell$ and $n$ coalesce.
A second, generalized mode survives, whose time dependence contains a term proportional to $t e^{-i\omega t}$, as follows from
\begin{equation}
\lim_{\epsilon\to0}\frac{e^{-i\omega_{+}t}-e^{-i\omega_{-}t}}{\omega_{+}-\omega_{-}}
=-it\,e^{-i\omega t},
\end{equation}
where $\omega$ is the common frequency at $\epsilon=0$.
This linear time dependence is the real-time manifestation of the Jordan block.

\subsection{Thermal correlators and real-time response}\label{sec:thermal}

At finite temperature, the dual CFT is in a thermal state, described holographically by a black hole.
Since the mapping (\ref{eq:field_map}) contains only constants, the eigenmodes $\psi_{\pm}$ remain decoupled.
Only their two-point functions change with the state.
For the planar nonrotating BTZ black hole ($d=2$, $x\in\mathbb{R}$) at temperature $T_{\text{H}}$, they take the standard finite-temperature form of a $(1+1)$-dimensional CFT \cite{SonStarinets2002JHEP}, in which the distance $\sepdist=|x+i\tau|$ is replaced by $r_{T}=|\sinh(\pi T_{\text{H}}(x+i\tau))|/(\pi T_{\text{H}})$.
The thermal two-point functions are thus the vacuum ones (\ref{eq: G_solution}) with $\sepdist$ replaced by $r_{T}$,
\begin{equation}
G(\tau,x)=r_{T}^{-2i\tilde{D}}G(1).\label{eq:thermal_euclidean}
\end{equation}
At large distance, $r_{T}\simeq e^{\pi T_{\text{H}}|x|}/(2\pi T_{\text{H}})$, so that the equal-time correlators decay as $G(0,x)\propto e^{-2\pi T_{\text{H}}|x|\,i\tilde{D}}$.
For $\epsilon>0$, they contain two correlation lengths, $\xi_{\pm}=1/(2\pi T_{\text{H}}\Delta_{\pm})$.
At the exceptional point, $\langle\mathscr{O}_{1}\mathscr{O}_{1}\rangle$ decays as $(a+b|x|)e^{-|x|/\xi}$ with constants $a$ and $b$ and $\xi=1/(2\pi T_{\text{H}}\Delta)$.
For $\epsilon<0$, the equal-time correlators oscillate in space with wave number $2\pi T_{\text{H}}\kappa$ and decay with the correlation length $1/(2\pi T_{\text{H}}\bar{\Delta})$.

\begin{figure}[t]
\centering
\includegraphics{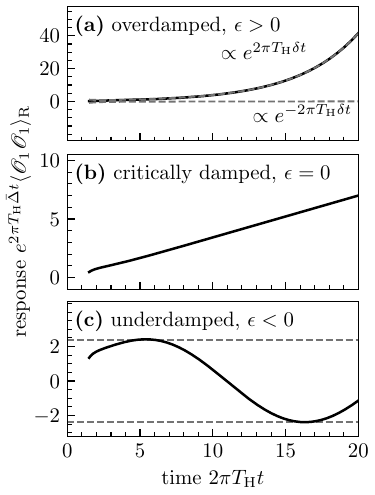}
\caption{Thermal retarded response $\langle\mathscr{O}_{1}\mathscr{O}_{1}\rangle_{\text{R}}$ at $x=0$, multiplied by $e^{2\pi T_{\text{H}}\bar{\Delta}t}$, against $2\pi T_{\text{H}}t$, in the BTZ black hole ($d=2$) with $m^{2}=-0.5$. \textbf{(a)} $\epsilon=0.1$: overdamped, two decay rates. The dashed curves are the contributions that decay as $e^{-2\pi T_{\text{H}}\Delta_{\pm}t}$ at late times. Multiplied by $e^{2\pi T_{\text{H}}\bar{\Delta}t}$, they scale as $e^{\mp2\pi T_{\text{H}}\delta t}$. \textbf{(b)} Exceptional point, $\epsilon=0$: critically damped, linear in $t$ at late times. \textbf{(c)} $\epsilon=-0.2$: underdamped, oscillation with frequency $2\pi T_{\text{H}}\kappa$. The dashed lines are its late-time envelope. Compare with the dependence on $\ln\sepdist$ in figure~\ref{fig:two_point}. See section~\ref{sec:thermal} for details.}
\label{fig:thermal}
\end{figure}

In real time, the analytic continuation $\tau\rightarrow it$ gives the retarded Green's functions, which describe the linear response.
Inside the light cone, $t>|x|$, they read
\begin{equation}
G_{\text{R}}(t,x)=-2\sin(\pi i\tilde{D})\,r_{\text{R}}^{-2i\tilde{D}}G(1),\qquad r_{\text{R}}=\frac{[\sinh(\pi T_{\text{H}}(t-x))\sinh(\pi T_{\text{H}}(t+x))]^{1/2}}{\pi T_{\text{H}}},\label{eq:thermal_retarded_full}
\end{equation}
and they vanish outside it.
We take the scaling dimensions to be non-integer.
At late times and fixed $x$, $r_{\text{R}}\simeq e^{\pi T_{\text{H}}t}/(2\pi T_{\text{H}})$, so that
\begin{equation}
G_{\text{R}}(t)\propto e^{-2\pi T_{\text{H}}t\,i\tilde{D}},\label{eq:thermal_retarded}
\end{equation}
so that the dilatation matrix generates damped dynamics.
As for a damped oscillator, the response is overdamped for $\epsilon>0$, underdamped for $\epsilon<0$, and critically damped at the exceptional point $\epsilon=0$, where it contains a term linear in time, $t\,e^{-2\pi T_{\text{H}}\Delta t}$ (figure~\ref{fig:thermal}).

In frequency space, the corresponding poles are the quasinormal frequencies \cite{SonStarinets2002JHEP,Birmingham2002PRL}
\begin{equation}
\omega=\pm k-4\pi iT_{\text{H}}\left(n+\frac{\Delta_{\pm}}{2}\right),\qquad n=0,1,2,\dots,\label{eq: btz_qnm}
\end{equation}
with momentum $k$, where the sign of $\pm k$ is independent of the label of $\Delta_{\pm}$.
Since $\mathrm{Re}\,\Delta_{\pm}>d/2$, all quasinormal modes are damped.
For $\epsilon>0$, the quasinormal frequencies of the eigenmodes $\psi_{+}$ and $\psi_{-}$ have the same real parts, $\pm k$, and different decay rates, $-\mathrm{Im}\,\omega=2\pi T_{\text{H}}(2n+\Delta_{\pm})$.
For $\epsilon<0$, they have the same decay rate, $2\pi T_{\text{H}}(2n+\bar{\Delta})$, and real parts $\pm k+2\pi T_{\text{H}}\kappa$ for $\Delta_{+}$ and $\pm k-2\pi T_{\text{H}}\kappa$ for $\Delta_{-}$.
The difference between the quasinormal frequencies of $\psi_{+}$ and $\psi_{-}$ at the same $k$, $n$, and sign of $\pm k$ is thus purely imaginary for $\epsilon>0$ and purely real for $\epsilon<0$.
In PT-symmetric coupled resonators, the difference between the two eigenfrequencies is instead real in the PT-unbroken phase and imaginary in the PT-broken phase \cite{Peng2014NatPhys}.
The roles are exchanged here because the scaling dimensions enter the quasinormal frequencies multiplied by $-i$.

At $\epsilon=0$, the quasinormal frequencies of $\psi_{+}$ and $\psi_{-}$ coalesce.
For $k\neq0$, the momentum-space retarded functions $G_{\text{R}}^{ab}(\omega,k)$ then have second-order poles in $G_{\text{R}}^{11}$, simple poles in $G_{\text{R}}^{12}$, and no poles in $G_{\text{R}}^{22}$.
At $k=0$, where the two signs in $\pm k$ give the same frequency, the orders of the poles of $G_{\text{R}}^{11}$ and $G_{\text{R}}^{12}$ increase by one.
A spatially uniform source therefore produces a response $\langle\mathscr{O}_{1}\mathscr{O}_{1}\rangle_{\text{R}}\propto t^{2}e^{-2\pi T_{\text{H}}\Delta t}$ at late times, compared with $t\,e^{-2\pi T_{\text{H}}\Delta t}$ for a single scalar operator.
Second-order poles also appear at exceptional points of quasinormal modes for a single scalar with mixed boundary conditions in the BTZ black hole \cite{WangYang2026JHEP,Long2026arXiv}.
Our model thus provides another route to realizing exceptional points in holography and studying their signatures in real-time dynamics.

\section{Two-point functions in hyperbolic electric circuits}\label{sec:circuit}

\begin{figure}[t]
\centering
\includegraphics{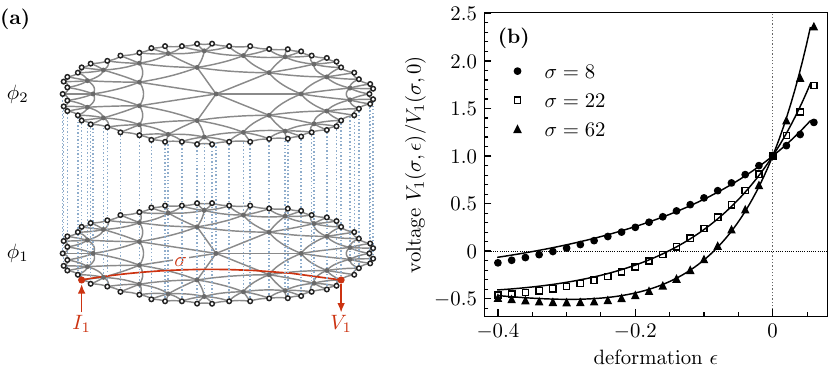}
\caption{Two coupled hyperbolic circuits tuned through the exceptional point. \textbf{(a)} Schematic illustration of two coupled hyperbolic circuits on the $\{3,7\}$ tessellation, one for each field $\phi_{a}$. Blue dotted lines join every node to its partner in the other circuit through $C_{b}$ and the INIC $C_{a}$. A current $I_{1}$ is injected at an edge node of circuit~1, and the voltage $V_{1}$ is measured at another edge node, at the invariant separation $\sigma$ of eq.~(\ref{eq:circuit_sigma}). \textbf{(b)} Voltage $V_{1}(\sigma,\epsilon)/V_{1}(\sigma,0)$ for a fixed current against $\epsilon$. Markers are circuit data, and lines are $\langle\mathscr{O}_{1}\mathscr{O}_{2}\rangle$ of eq.~(\ref{eq:PT_AdS_2pt_2}), with two constants taken from the circuits at the exceptional point. $V_{1}$ changes continuously through the exceptional point (dotted line) and changes sign at an $\epsilon<0$ that approaches the exceptional point as $\sigma$ grows. Parameters: $1625$ nodes per circuit, $m^{2}=0.25$, and $C_{b}=C$; the INIC capacitance $C_{a}$ sets $\epsilon$ through eq.~(\ref{eq:circuit_eps}). See section~\ref{subsec:circuit_two} for details.}
\label{fig:circuit_measurement}
\end{figure}

In this section, we show that two coupled electric circuits on a hyperbolic lattice realize the deformed dipole model (\ref{eq:the_action}), including its exceptional point.
Electric circuits on hyperbolic lattices realize the equation of motion of a single scalar field in Euclidean AdS spacetime \cite{Lenggenhager2022NC,Chen2023arXiv,Dey2024PRL}.
In such a circuit, the voltage at one edge node, in response to an electric current injected at another edge node, reproduces the two-point function of the dual CFT, which is a single power law with a real scaling dimension \cite{Asaduzzaman2020PRD,Chen2023arXiv}.
Ref.~\cite{Chen2023arXiv} measured this power law and confirmed the relation between the scaling dimension and the squared mass.
To realize two coupled fields, we join two such circuits at every node by a capacitor and by a non-reciprocal coupling, an operational-amplifier circuit that acts with different strengths in the two directions \cite{Helbig2020NP} (figure~\ref{fig:circuit_measurement}(a)).
The coupled circuits then have the exceptional point of the scaling dimensions.
In the following, we summarize how a single circuit realizes a scalar field in AdS (section~\ref{subsec:circuit_one}), and then couple two circuits and read the two-point functions from the voltages at their edge nodes (section~\ref{subsec:circuit_two}).

\subsection{A scalar field in a hyperbolic circuit}\label{subsec:circuit_one}

The circuits of refs.~\cite{Lenggenhager2022NC,Chen2023arXiv,Dey2024PRL} start from a regular tessellation $\{p,q\}$ of the hyperbolic plane, which covers the plane with regular $p$-gons, $q$ of which meet at every vertex \cite{Asaduzzaman2020PRD}.
The tessellation is drawn on the \Poincare disk $|\chi|<1$, with the complex coordinate $\chi$ and the metric $4|d\chi|^{2}/(1-|\chi|^{2})^{2}$, which is Euclidean AdS$_{2}$ with unit AdS radius.
A circuit keeps the vertices that a path of at most a given number of edges connects to a vertex at the center of the disk, so that the kept vertices form rings around the center.
The vertices of the outermost ring, the edge nodes, approximate the boundary of AdS$_{2}$ \cite{Asaduzzaman2020PRD,Chen2023arXiv}.
Each vertex becomes a node, each edge becomes a capacitor $C$ between two nodes, and an additional element connects every node to ground, the reference at zero voltage \cite{Lenggenhager2022NC,Chen2023arXiv}.
Currents injected at a fixed angular frequency $\omega$ make every current and every voltage oscillate at this frequency.
Each of them is described by its complex amplitude: the voltage of node $j$ is $\mathrm{Re}(V_{j}e^{i\omega t})$, where the modulus of $V_{j}$ is the amplitude of the oscillation and its argument is the phase.
Charge conservation at every node then gives a linear relation between the injected currents $I_{j}$ and the voltages $V_{j}$ \cite{Lee2018CommPhys,Lenggenhager2022NC},
\begin{equation}
I=i\omega C\,MV,\qquad M=-A+q+y,\label{eq:circuit_node}
\end{equation}
where $A$ is the adjacency matrix of the tessellation and the element to ground gives the term $y$ (appendix~\ref{app:circuit}).
The matrix $M$ is the circuit Laplacian in units of $i\omega C$, and its inverse is the circuit Green's function \cite{Lee2018CommPhys,Lenggenhager2022NC}.

For a voltage that varies slowly over one edge, $-A+q$ becomes $-s\nabla_{\mu}\nabla^{\mu}$, with $\nabla_{\mu}\nabla^{\mu}$ the Laplacian of the hyperbolic plane, $s=qh^{2}$, and $h$ the value of $|\chi|$ at the nearest neighbors of the vertex at the center \cite{Lenggenhager2022NC} (appendix~\ref{app:circuit}).
Equation~(\ref{eq:circuit_node}) then becomes the Klein--Gordon equation of a scalar field in Euclidean AdS$_{2}$ \cite{Chen2023arXiv,Dey2024PRL},
\begin{equation}
s\left(-\nabla_{\mu}\nabla^{\mu}+m^{2}\right)V=\frac{I}{i\omega C},\qquad m^{2}=\frac{y}{s},\label{eq:circuit_KG}
\end{equation}
where the voltage $V_{j}$ plays the role of the field at the vertex $j$, and the injected current $I_{j}$ the role of a source at this vertex.
The voltage at one node per current injected at another node is therefore proportional to the bulk-to-bulk propagator between the two nodes.
Between two points near the boundary, this propagator decays as $e^{-\Delta L}$, with $L$ the geodesic distance between the two points and $\Delta(\Delta-1)=m^{2}$, which is eq.~(\ref{eq:Delta_pm}) at $d=1$ and $\epsilon=0$ \cite{Asaduzzaman2020PRD,Chen2023arXiv}.
For two points $\chi$ and $\chi^{\prime}$ on the disk, $\cosh L=1+2\sigma^{2}$, with the invariant separation
\begin{equation}
\sigma=\frac{|\chi-\chi^{\prime}|}{[(1-|\chi|^{2})(1-|\chi^{\prime}|^{2})]^{1/2}}.\label{eq:circuit_sigma}
\end{equation}
Like $L$, $\sigma$ does not change under the isometries of the disk.
At large $\sigma$, $e^{-\Delta L}\simeq(2\sigma)^{-2\Delta}$.
The voltage at one edge node per current injected at another edge node therefore decays as $\sigma^{-2\Delta}$, which is the two-point function of the dual CFT, with $\sigma$ in the role of the separation $\sepdist$ of the two boundary points \cite{Asaduzzaman2020PRD,Chen2023arXiv}.

\begin{figure}[t]
\centering
\includegraphics{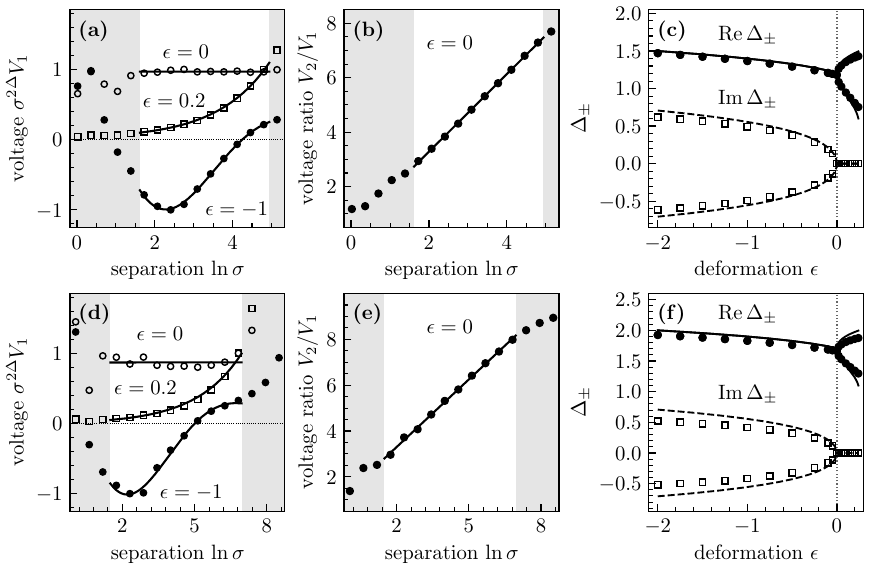}
\caption{Edge voltages of the two coupled circuits, against the invariant separation $\sigma$ of eq.~(\ref{eq:circuit_sigma}), for the $\{3,7\}$ tessellation ($d=1$, top row) and the $\{4,3,5\}$ honeycomb ($d=2$, bottom row). Markers are circuit data. A current $I_{1}$ is injected at an edge node of circuit~1, as in figure~\ref{fig:circuit_measurement}(a), and $V_{a}$ is the voltage at another edge node of circuit~$a$, so that $V_{a}$ is proportional to the voltage response $G_{a1}$, with $G_{ab}=i\omega C\,V_{a}/I_{b}$. \textbf{(a,d)} $V_{1}$, in arbitrary units, for $\epsilon=0.2$ (open squares), $\epsilon=0$ (open circles), and $\epsilon=-1$ (filled circles), multiplied by $\sigma^{2\Delta}$, with $\Delta$ the scaling dimension of a single circuit with squared mass $m^{2}$. Each of the lines is fitted separately to its own data, with two free parameters (one at $\epsilon=0$): the coefficients of $\sigma^{-2\Delta_{+}}$ and $\sigma^{-2\Delta_{-}}$, the two terms of $\langle\mathscr{O}_{1}\mathscr{O}_{2}\rangle$ in eq.~(\ref{eq:PT_AdS_2pt_2}), with the scaling dimensions of the circuits in (c) and (f) as exponents. At $\epsilon=0$, $V_{1}$ reduces to the single power law $\sigma^{-2\Delta}$, as $\langle\mathscr{O}_{1}\mathscr{O}_{2}\rangle$ does. The logarithm then appears in $V_{2}$, shown in (b) and (e). $V_{1}$ stays positive for $\epsilon\geq0$ and changes sign for $\epsilon=-1$. Figure~\ref{fig:circuit_measurement}(b) shows $V_{1}$ of the top row at three fixed separations as $\epsilon$ is tuned. \textbf{(b,e)} Voltage ratio $V_{2}/V_{1}=G_{21}/G_{11}$ at $\epsilon=0$, with a linear fit in $\ln\sigma$ (line). The ratio is linear in $\ln\sigma$, which is the logarithm of the LCFT. \textbf{(c,f)} Scaling dimensions of the circuits (markers), real parts (filled circles) and imaginary parts (open squares), obtained from the slopes of the logarithms of the two eigenvalues of $G(\sigma)G(\sigma_{0})^{-1}$ against $\ln\sigma$. The lines are the analytic dimensions $\Delta_{\pm}$ of eq.~(\ref{eq:Delta_pm}) at the $d$ and $m^{2}$ of each row, without free parameters: real parts (solid) and imaginary parts (dashed). The dotted lines mark the exceptional point. In (a), (b), (d), and (e), the shaded ranges lie outside the window of the fits: at short separations, the lattice spacing matters, and at the largest separations, the finite size of the circuit matters. Top row: $1625$ nodes per circuit, $m^{2}=0.25$, and $C_{b}=C$. Bottom row: $3.6\times10^{6}$ nodes per circuit, $m^{2}=-0.5$, and $C_{b}=2C$. Here, $C$ is the capacitance between neighboring nodes, $C_{b}$ the capacitor between the two circuits, and the INIC capacitance $C_{a}$ sets $\epsilon$ through eq.~(\ref{eq:circuit_eps}). See section~\ref{subsec:circuit_two} for details.}
\label{fig:circuit}
\end{figure}

\subsection{Two coupled circuits and their edge voltages}\label{subsec:circuit_two}

To realize the model (\ref{eq:the_action}), we take two copies of the circuit, $a=1,2$, and identify the voltage of circuit $a$ with the field $\phi_{a}$.
In the equations of motion (\ref{eq:PT_modified_EOM}), $\phi_{2}$ enters the equation of $\phi_{1}$ with the coefficient $\epsilon$, and $\phi_{1}$ enters the equation of $\phi_{2}$ with the coefficient $1$.
For $\epsilon\leq0$, the coupling between the two circuits must therefore act differently in the two directions, as a negative impedance converter with current inversion (INIC), an operational-amplifier circuit, does \cite{Helbig2020NP}.
Between two nodes, the INIC acts as a capacitance $C_{a}$ seen from one node and as a capacitance $-C_{a}$ seen from the other \cite{Helbig2020NP}.
At every node, we join the two circuits by a capacitor $C_{b}$ and, in parallel, an INIC $C_{a}$, which gives (appendix~\ref{app:circuit})
\begin{equation}
\epsilon=\frac{C_{b}^{2}-C_{a}^{2}}{s^{2}C^{2}}.\label{eq:circuit_eps}
\end{equation}
Thus $\epsilon$ is positive for $C_{a}<C_{b}$ and negative for $C_{a}>C_{b}$.
For $C_{a}=C_{b}$, the first circuit drives the second but not the reverse, and the coupled circuits realize the dipole model.
Notably, the coupled circuits have the exceptional point on the lattice itself.
The coupling acts at every node through the same $2\times2$ matrix on the circuit index, so that it commutes with the circuit Laplacian $M$, which acts on the node index.
Diagonalizing the $2\times2$ matrix therefore decouples the two circuits into the eigenmodes $\psi_{\pm}$, with squared masses $m^{2}\pm\sqrt{\epsilon}$, on any finite tessellation.
At $\epsilon=0$, the $2\times2$ matrix is a rank-2 Jordan block, as is the mixing matrix in eq.~(\ref{eq:mixing_matrix}).

The coupled circuits measure the two-point functions directly.
We inject an alternating current $I_{b}$ at an edge node of circuit $b$ and measure the voltage $V_{a}$ at another edge node of circuit $a$ (figure~\ref{fig:circuit_measurement}(a)), as ref.~\cite{Chen2023arXiv} did for a single circuit.
As for a single circuit, the dependence of the response $G_{ab}=i\omega C\,V_{a}/I_{b}$ on $\sigma$ gives the two-point functions.
For every $\epsilon$, $G_{11}=G_{22}$ and $G_{12}/G_{21}=(C_{b}-C_{a})/(C_{b}+C_{a})$ (appendix~\ref{app:circuit}).
Up to a rescaling of the voltages and currents of the second circuit (appendix~\ref{app:circuit}), $G_{11}$, $G_{21}$, and $G_{12}$ therefore have the forms of $\langle\mathscr{O}_{1}\mathscr{O}_{2}\rangle$, $\langle\mathscr{O}_{1}\mathscr{O}_{1}\rangle$, and $\langle\mathscr{O}_{2}\mathscr{O}_{2}\rangle$ in eqs.~(\ref{eq:PT_AdS_2pt_1})--(\ref{eq:PT_AdS_2pt_3}), with normalization constants set by the lattice.

The edge voltages show the three regimes of the two-point functions [figure~\ref{fig:circuit}(a--c) for $d=1$].
In figure~\ref{fig:circuit}(a), $G_{11}$ stays positive for $\epsilon=0.2$ and $\epsilon=0$.
For $\epsilon=-1$, it oscillates in $\ln\sigma$ and changes sign, as $\langle\mathscr{O}_{1}\mathscr{O}_{2}\rangle$ in eq.~(\ref{eq:PT_AdS_2pt_2}).
Since $G_{11}\geq0$ for $\epsilon\geq0$ (appendix~\ref{app:circuit}), a sign change of the edge voltage occurs only in the PT-broken phase.
At the exceptional point, $G_{21}/G_{11}$ is linear in $\ln\sigma$ [figure~\ref{fig:circuit}(b)], which is the logarithm of the LCFT.
Finally, as in section~\ref{sec:CFT_calculation}, the eigenvalues of $G(\sigma)G(\sigma_{0})^{-1}$ decay as $(\sigma/\sigma_{0})^{-2\Delta_{\pm}}$ and give the two scaling dimensions, which coalesce at the exceptional point as the continuum values do, up to lattice corrections (appendix~\ref{app:circuit}) [figure~\ref{fig:circuit}(c)].
Since the $\{3,7\}$ tessellation has a one-dimensional boundary, these circuits test the scaling dimensions and the form of the two-point functions of section~\ref{sec:AdS_calculation} at $d=1$.
The derivation from the conformal Ward identities in section~\ref{sec:CFT_calculation}, however, assumes $d\geq2$.
For $d=2$, we build the same circuits on the order-5 cubical honeycomb $\{4,3,5\}$ of hyperbolic 3-space \cite{Asaduzzaman2020PRD} (appendix~\ref{app:circuit}).
Their edge voltages show the same three regimes [figure~\ref{fig:circuit}(d--f)], and thus test the two-point functions where both derivations apply.

In an experiment, the nodes stay fixed, and $C_{a}$ tunes $\epsilon$ (figure~\ref{fig:circuit_measurement}(b)).
At a fixed separation, the voltage $V_{1}$ then changes continuously through the exceptional point, as the two-point functions do.
It changes sign at an $\epsilon<0$ that approaches the exceptional point as $\sigma$ grows.
In the PT-broken phase, ideal circuits have growing modes, the counterpart of the growing modes in global AdS (section~\ref{sec:realtime}).
The simulations assume ideal components; stabilizing these modes in an experiment is not studied here.

\section{Conclusion}\label{sec: Discussion}

We demonstrate that, in a deformed dipole model in Euclidean AdS spacetime, the holographic two-point functions continue those of an LCFT through its exceptional point.
The two scaling dimensions are real on one side of the exceptional point and complex conjugate on the other.
Correspondingly, the holographic two-point functions are linear combinations of two power laws in the former case and exhibit log-periodic oscillations in the latter.
Approaching the exceptional point, the two power laws merge into the logarithmic two-point functions of the LCFT dual to the dipole model.
The same two-point functions follow from the conformal Ward identities of the dual CFT, up to two normalization constants.
Since every real dilatation matrix of a pair of scalar primary operators that is not proportional to the identity takes the same form in a suitable basis, this form describes all real deformations of the exceptional point.
In PT-symmetric theories with a dilatation matrix of this form, these two-point functions therefore characterize the PT-unbroken phase, the PT-broken phase, and the exceptional point separating them.
For the present model, this identification holds if time reversal is preserved by the boundary conditions, the quantization, and the vacuum.
At finite temperature, the same dilatation matrix governs the linear response in the BTZ black hole.
The quasinormal frequencies associated with the two scaling dimensions then differ by a purely imaginary amount for real dimensions and by a purely real amount for complex-conjugate dimensions.
The response changes from overdamped to underdamped, with critical damping at the exceptional point.
The two-point functions across the exceptional point are realized in two hyperbolic electric circuits with a non-reciprocal coupling.
These circuits realize the model with an exact exceptional point on any finite lattice.
Their voltage response between two edge nodes changes sign only in the PT-broken phase.
At the exceptional point, the response between the two circuits acquires the logarithm of the LCFT.

This study can be extended in several directions.
First, higher-derivative gravity theories host similar structures.
For example, six-derivative gravity in AdS$_{3}$ has a curve in parameter space on which the squared masses of two massive gravitons coalesce \cite{Bergshoeff2012PRD}, and conformal gravity has a Jordan-block structure in its PT-symmetric quantization \cite{BenderMannheim2008PRD,Mannheim2012FP}.
Analogous deformations of these theories and their holographic two-point functions are natural next steps.
Second, Jordan-block structures are also proposed in flat-space holography \cite{Fiorucci2024PRD,Bagchi2026EPJC}.
A natural question is whether similar deformations, and the corresponding real-to-complex transitions, can be realized in celestial and Carrollian conformal field theories.
Third, the theory considered here is non-unitary with the conventional inner product, as is typical for LCFTs.
Whether the regime with real scaling dimensions admits a quantization with a positive inner product, as for the fourth-order oscillator \cite{BenderMannheim2008PRL}, is left for future work.
Finally, our thermal results describe the linear response at a fixed value of $\epsilon$.
A quench of $\epsilon$ across the exceptional point, analogous to the quenches of non-Hermitian couplings in ref.~\cite{Sergio_2023_SciPost}, deserves further study. We believe our work will stimulate further investigations of non-Hermitian physics in holography.

\acknowledgments
The authors used ChatGPT (OpenAI, GPT-6.1 sol) and Claude (Anthropic, Claude Opus 5.5) during the preparation of this manuscript to assist with code optimization and to improve the readability of the manuscript. The authors assume full responsibility for the scientific content, accuracy, and integrity of this paper.

This work was supported by the NKRDPC (Grant No.~2022YFA1405304), NSFC (Grant No.~12275089), and Guangdong Provincial Quantum Science Strategic Initiative (Grant Nos.~GDZX2304002 and~GDZX2404001).

\appendix

\section{Generality of the deformation}\label{app:generality}

Here, we establish the generality of the action (\ref{eq:the_action}) within real quadratic two-scalar theories that admit a mass-matrix exceptional point, and of the dilatation matrix (\ref{eq:perturbed_D}) on a closed pair of scalar primary operators.

\subsection{Two-field actions in the bulk}

Consider two real scalar fields $\Phi=(\phi_{1},\phi_{2})^{T}$ with quadratic action
\begin{equation}
S=-\frac{1}{2}\int d^{d+1}x\sqrt{g}\left[K_{ab}\nabla_{\mu}\phi_{a}\nabla^{\mu}\phi_{b}+V_{ab}\phi_{a}\phi_{b}\right],\label{eq:general_action}
\end{equation}
where $K$ and $V$ are constant real symmetric $2\times2$ matrices and $K$ is invertible.
The equations of motion read $\nabla_{\mu}\nabla^{\mu}\Phi=K^{-1}V\Phi$, so that the squared masses are the eigenvalues of $K^{-1}V$.
If $K$ is definite, a real field redefinition brings it to $\pm I$, making the transformed mass matrix real symmetric and hence diagonalizable with real eigenvalues.
An exceptional point of the mass matrix therefore requires indefinite $K$, with one negative kinetic direction.

For indefinite $K$, a real constant field redefinition gives the kinetic term $2\nabla_{\mu}\phi_{1}\nabla^{\mu}\phi_{2}$, with potential $V_{11}\phi_{1}^{2}+2V_{12}\phi_{1}\phi_{2}+V_{22}\phi_{2}^{2}$.
The rescaling $\phi_{1}\rightarrow\lambda\phi_{1}$, $\phi_{2}\rightarrow\phi_{2}/\lambda$ preserves the kinetic term and sends $V_{11}\rightarrow\lambda^{2}V_{11}$ and $V_{22}\rightarrow V_{22}/\lambda^{2}$, leaving $V_{12}$ and $V_{11}V_{22}$ unchanged.
Exchanging the fields also preserves the kinetic term.
If $V_{11}=V_{22}=0$, the mass matrix is proportional to the identity and has no Jordan block.
Otherwise, exchange and rescale the fields to obtain $|V_{11}|=1$.
For $V_{11}=-1$, reversing the overall sign of the action and sending $\phi_{1}\rightarrow-\phi_{1}$ gives $V_{11}=1$ and preserves the kinetic term and $V_{12}$.
The action then takes the form (\ref{eq:the_action}) with
\begin{equation}
m^{2}=V_{12},\qquad\epsilon=V_{11}V_{22},
\end{equation}
up to an overall sign, which reverses the holographic two-point functions.

\subsection{Dilatation matrices on the CFT side}

Consider a pair of real scalar operators closed under dilatations, with a real dilatation matrix
\begin{equation}
iD|\mathscr{O}_{a}\rangle=(i\tilde{D})_{a}{}^{b}|\mathscr{O}_{b}\rangle,\qquad i\tilde{D}=\left(\begin{array}{cc}
\Delta_{11} & \Delta_{12}\\
\Delta_{21} & \Delta_{22}
\end{array}\right).
\end{equation}
If $\Delta_{12}\neq0$, we define the new operators
\begin{equation}
\mathscr{O}_{1}^{\prime}=\mathscr{O}_{1},\qquad\mathscr{O}_{2}^{\prime}=\frac{\Delta_{11}-\Delta_{22}}{2}\mathscr{O}_{1}+\Delta_{12}\mathscr{O}_{2}.
\end{equation}
Their dilatation action is
\begin{equation}
iD|\mathscr{O}_{1}^{\prime}\rangle=\bar{\Delta}|\mathscr{O}_{1}^{\prime}\rangle+|\mathscr{O}_{2}^{\prime}\rangle,\qquad iD|\mathscr{O}_{2}^{\prime}\rangle=\eta|\mathscr{O}_{1}^{\prime}\rangle+\bar{\Delta}|\mathscr{O}_{2}^{\prime}\rangle,
\end{equation}
with
\begin{equation}
\bar{\Delta}=\frac{\Delta_{11}+\Delta_{22}}{2},\qquad\eta=\frac{(\Delta_{11}-\Delta_{22})^{2}}{4}+\Delta_{12}\Delta_{21},
\end{equation}
i.e., the matrix takes the form (\ref{eq:perturbed_D}).
If $\Delta_{12}=0$ but $\Delta_{21}\neq0$, exchange the operators.
For a diagonal matrix with unequal entries, choose $\mathscr{O}_{1}^{\prime}=\mathscr{O}_{1}+\mathscr{O}_{2}$ and $|\mathscr{O}_{2}^{\prime}\rangle=(iD-\bar{\Delta})|\mathscr{O}_{1}^{\prime}\rangle$.

\section{Holographic renormalization of the dipole model}\label{app:dipole}

Here, we derive the holographic two-point functions of the dipole model directly at $\epsilon=0$, which provides a check of the $\epsilon\rightarrow0$ limit in section~\ref{sec:AdS_calculation}.
For the dipole model, $S_{\text{sd}}$, the equations of motion are 
\begin{equation}
(\nabla_{\mu}\nabla^{\mu}-m^{2})\phi_{1}=0,\quad(\nabla_{\mu}\nabla^{\mu}-m^{2})\phi_{2}=\phi_{1}.\label{eq: dipole_EOMs}
\end{equation}
These equations of motion can be solved using the Green's function method, where we construct the bulk-to-boundary propagator $K_{ij}$ by 
\begin{align}
\phi_{1}(z,x^{i}) & =\int d^{d}x^{\prime}K_{11}(z,x^{i};0,x^{i\prime})\phi_{1(0)}(x^{i\prime}),\\
\phi_{2}(z,x^{i}) & =\int d^{d}x^{\prime}\left(K_{12}(z,x^{i};0,x^{i\prime})\phi_{1(0)}(x^{i\prime})+K_{22}(z,x^{i};0,x^{i\prime})\phi_{2(0)}(x^{i\prime})\right).
\end{align}
Now, we only need to solve the following equations for the propagators, 
\begin{equation}
\begin{array}{c}
(\nabla_{\mu}\nabla^{\mu}-m^{2})K_{11}=0,\\
(\nabla_{\mu}\nabla^{\mu}-m^{2})K_{12}=K_{11},\quad(\nabla_{\mu}\nabla^{\mu}-m^{2})K_{22}=0.
\end{array}
\end{equation}
The propagators of the Klein--Gordon mode \cite{witten1998ATMP} and of the logarithmic mode, $K_{12}$ \cite{GHEZELBASH1999IJMPA,KOGAN1999PLB,Bergshoeff2012JHEP}, are
\begin{align}
K_{11}(z,x^{i};0,x^{i\prime})= & K_{22}(z,x^{i};0,x^{i\prime})=\frac{1}{N_{\Delta}}\frac{z^{\Delta}}{(z^{2}+\left|x-x^{\prime}\right|^{2})^{\Delta}},\\
K_{12}(z,x^{i};0,x^{i\prime})= & \frac{1}{2\Delta-d}\partial_{\Delta}K_{11}(z,x^{i};0,x^{i\prime})\nonumber \\
= & \frac{1}{(2\Delta-d)}\frac{1}{N_{\Delta}}\left(\ln\left(\frac{z}{z^{2}+\left|x-x^{\prime}\right|^{2}}\right)-\partial_{\Delta}\ln N_{\Delta}\right)\frac{z^{\Delta}}{(z^{2}+\left|x-x^{\prime}\right|^{2})^{\Delta}},
\end{align}
where $\Delta=(d+\sqrt{d^{2}+4m^{2}})/2$.
Note that the normalization $1/N_{\Delta}$ of $K_{11}$ and $K_{22}$ is fixed to ensure the boundary behaviors 
\begin{align}
\lim_{z\rightarrow0}\phi_{1}(z,x^{i})= & \lim_{z\rightarrow0}z^{d-\Delta}\phi_{1(0)}(x^{i}),\\
\lim_{z\rightarrow0}\phi_{2}(z,x^{i})= & \lim_{z\rightarrow0}\int d^{d}x^{\prime}K_{12}(z,x^{i};0,x^{i\prime})\phi_{1(0)}(x^{i\prime})+\lim_{z\rightarrow0}z^{d-\Delta}\phi_{2(0)}(x^{i}),
\end{align}
where we used (\ref{eq: delta_function}).
The on-shell action is given by (\ref{eq:boundary_action}).
We first evaluate it on the cutoff surface $z=\zeta$, 
\begin{equation}
S_{\text{reg}}=\frac{1}{2}\int_{z=\zeta}d^{d}xz^{-d}\left(\phi_{1}z\partial_{z}\phi_{2}+\phi_{2}z\partial_{z}\phi_{1}\right)=\frac{1}{2}\int_{z=\zeta}d^{d}xz^{-d}z\partial_{z}\left(\phi_{1}\phi_{2}\right).
\end{equation}
Near the boundary, the bulk-to-boundary propagator behaves as 
\begin{equation}
\frac{1}{N_{\Delta}}\frac{z^{\Delta}}{(z^{2}+\left|x-x^{\prime}\right|^{2})^{\Delta}}=z^{d-\Delta}\delta^{d}(x-x^{\prime})+\frac{1}{N_{\Delta}}\frac{z^{\Delta}}{\left|x-x^{\prime}\right|^{2\Delta}}+\cdots,\label{eq: near_boundary_propagator}
\end{equation}
where the first term reproduces eq.~(\ref{eq: delta_function}), the second term gives the behavior at $x\neq x^{\prime}$, and the dots denote terms of higher order in $z$.
Both terms are needed.
Since $z^{-d}\phi_{i}$ grows as $z^{-\Delta}$, the second term contributes to the on-shell action at finite order.
Differentiating (\ref{eq: near_boundary_propagator}) with respect to $\Delta$, we find 
\begin{align}
& \partial_{\Delta}\left(\frac{1}{N_{\Delta}}\frac{z^{\Delta}}{(z^{2}+\left|x-x^{\prime}\right|^{2})^{\Delta}}\right) \nonumber\\
=& -\ln z\,z^{d-\Delta}\delta^{d}(x-x^{\prime})+\frac{\ln z-\ln\left|x-x^{\prime}\right|^{2}-\partial_{\Delta}\ln N_{\Delta}}{N_{\Delta}}\frac{z^{\Delta}}{\left|x-x^{\prime}\right|^{2\Delta}}+\cdots.
\end{align}
The fields then read 
\begin{align}
\phi_{1}= & z^{d-\Delta}\phi_{1(0)}+z^{\Delta}B_{1}+\cdots,\\
\phi_{2}= & z^{d-\Delta}\left(\phi_{2(0)}-\frac{\ln z}{2\Delta-d}\phi_{1(0)}\right)+z^{\Delta}\left(B_{2}+\frac{B_{1}\ln z-C_{1}-B_{1}\partial_{\Delta}\ln N_{\Delta}}{2\Delta-d}\right)+\cdots,
\end{align}
where $B_{i}(x)\equiv\frac{1}{N_{\Delta}}\int d^{d}x^{\prime}\frac{\phi_{i(0)}(x^{\prime})}{\left|x-x^{\prime}\right|^{2\Delta}}$, and $C_{1}(x)\equiv\frac{1}{N_{\Delta}}\int d^{d}x^{\prime}\frac{\phi_{1(0)}(x^{\prime})\ln\left|x-x^{\prime}\right|^{2}}{\left|x-x^{\prime}\right|^{2\Delta}}$.
Plugging them into $S_{\text{reg}}$, we find 
\begin{align}
S_{\text{reg}}= & \frac{1}{2}\int_{z=\zeta}d^{d}x\,z^{d-2\Delta}\left(2(d-\Delta)\phi_{1(0)}\left(\phi_{2(0)}-\frac{\ln z}{2\Delta-d}\phi_{1(0)}\right)-\frac{\phi_{1(0)}^{2}}{2\Delta-d}\right)\nonumber \\
 & +\frac{d}{2}\int d^{d}x\left(\phi_{1(0)}\left(B_{2}-\frac{C_{1}+B_{1}\partial_{\Delta}\ln N_{\Delta}}{2\Delta-d}\right)+B_{1}\phi_{2(0)}\right)+\cdots.
\end{align}
The first term with $z^{d-2\Delta}$ diverges as $\zeta\rightarrow0$.
It contains the sources only at the same point $x$ and is thus a contact term, which does not contribute to the two-point functions at separated points.
Following holographic renormalization \cite{Kostas2002CQG}, it is canceled by a counterterm \cite{Bergshoeff2011JHEP}, 
\begin{equation}
S_{\text{ct}}=\int_{z=\zeta}d^{d}x\sqrt{\gamma}\left(\frac{\phi_{1}^{2}}{2(2\Delta-d)}-(d-\Delta)\phi_{1}\phi_{2}\right),
\end{equation}
whose finite part is $\int d^{d}x\left(\frac{\phi_{1(0)}B_{1}}{2\Delta-d}-(d-\Delta)\left(\phi_{1(0)}\left(B_{2}-\frac{C_{1}+B_{1}\partial_{\Delta}\ln N_{\Delta}}{2\Delta-d}\right)+B_{1}\phi_{2(0)}\right)\right)$.
For $2\Delta-d\geq2$, counterterms with derivatives along the boundary remove further divergent contact terms \cite{Kostas2002CQG,Bergshoeff2011JHEP}.
These counterterms do not contribute at separated points.
We assume $\Delta-d/2\notin\mathbb{Z}$.
Otherwise, logarithmic counterterms are also needed \cite{Kostas2002CQG}, which do not contribute at separated points either.
The renormalized on-shell action, $S_{\text{ren}}\equiv\lim_{\zeta\rightarrow0}(S_{\text{reg}}+S_{\text{ct}})$, is then finite and reads 
\begin{align}
S_{\text{ren}}= & \int d^{d}x\left(\frac{2\Delta-d}{2}\left(\phi_{1(0)}\left(B_{2}-\frac{C_{1}+B_{1}\partial_{\Delta}\ln N_{\Delta}}{2\Delta-d}\right)+B_{1}\phi_{2(0)}\right)+\frac{\phi_{1(0)}B_{1}}{2\Delta-d}\right)\nonumber \\
= & \int d^{d}xd^{d}x^{\prime}\frac{\phi_{1(0)}(x)\phi_{2(0)}(x^{\prime})}{\left|x-x^{\prime}\right|^{2\Delta}}2\tilde{A}-\int d^{d}xd^{d}x^{\prime}\frac{\phi_{1(0)}(x)\phi_{1(0)}(x^{\prime})}{\left|x-x^{\prime}\right|^{2\Delta}}\frac{\tilde{A}\ln\left|x-x^{\prime}\right|^{2}-\partial_{\Delta}\tilde{A}}{2\Delta-d},\label{eq: diverging_onshell_action}
\end{align}
where $A\equiv\frac{2\Delta-d}{2}$ and $\tilde{A}\equiv A/N_{\Delta}=\frac{2\Delta-d}{2N_{\Delta}}$.
With the coupling (\ref{eq:GKPW}), the \emph{connected} two-point functions at separated points are 
\begin{align}
\langle\mathscr{O}_{1}(x_{1})\mathscr{O}_{1}(x_{2})\rangle= & \frac{\delta}{\delta\phi_{1(0)}(x_{1})}\frac{\delta}{\delta\phi_{1(0)}(x_{2})}\ln\left\langle \exp\int d^{d}x(\phi_{1(0)}\mathscr{O}_{1}+\beta\phi_{2(0)}\mathscr{O}_{2})\right\rangle _{\text{CFT}}\nonumber \\
= & -\frac{\delta}{\delta\phi_{1(0)}(x_{1})}\frac{\delta}{\delta\phi_{1(0)}(x_{2})}S_{\text{ren}}|_{\phi_{1}(\partial\text{AdS})=\phi_{1(0)},\phi_{2}(\partial\text{AdS})=\phi_{2(0)}}\nonumber \\
= & -\frac{1}{2\Delta-d}\partial_{\Delta}\left(\frac{2\tilde{A}}{\left|x_{1}-x_{2}\right|^{2\Delta}}\right)\nonumber \\
= & \frac{2\tilde{A}}{2\Delta-d}\frac{2\ln\left|x_{1}-x_{2}\right|}{\left|x_{1}-x_{2}\right|^{2\Delta}}-\frac{2\partial_{\Delta}\tilde{A}}{2\Delta-d}\frac{1}{\left|x_{1}-x_{2}\right|^{2\Delta}},\label{eq:diverging_2ptFunction_1}\\
\beta\langle\mathscr{O}_{1}(x_{1})\mathscr{O}_{2}(x_{2})\rangle= & -\frac{\delta}{\delta\phi_{1(0)}(x_{1})}\frac{\delta}{\delta\phi_{2(0)}(x_{2})}S_{\text{ren}}|_{\phi_{1}(\partial\text{AdS})=\phi_{1(0)},\phi_{2}(\partial\text{AdS})=\phi_{2(0)}}\nonumber \\
= & -\frac{2\tilde{A}}{\left|x_{1}-x_{2}\right|^{2\Delta}}.\label{eq:diverging_2ptFunction_2}
\end{align}
The counterterm used in holographic renormalization \cite{Kostas2002CQG} removes the divergent contact terms, including a $\ln z$ divergence.
Its finite part fixes the normalization of the two-point functions at separated points.
The logarithmic two-point function (\ref{eq:diverging_2ptFunction_1}) is the $\Delta$-derivative of $\beta\langle\mathscr{O}_{1}\mathscr{O}_{2}\rangle$ in (\ref{eq:diverging_2ptFunction_2}), divided by $2\Delta-d$, which is the structure found in ref.~\cite{GHEZELBASH1999IJMPA}.
Holographic renormalization of these models is also carried out in refs.~\cite{Bergshoeff2011JHEP,Bergshoeff2012JHEP}.
With $2\Delta-d=\sqrt{d^{2}+4m^{2}}$, these two-point functions coincide with the $\epsilon\rightarrow0$ limit (\ref{eq:finite_2ptFunction_1})--(\ref{eq:finite_2ptFunction_2}) in section~\ref{sec:AdS_calculation}.

\section{Holographic renormalization of the deformed dipole model}\label{app: epsilon}

The on-shell action (\ref{eq:boundary_action}) on the cutoff surface for (\ref{eq:the_action}) takes the same form as in appendix~\ref{app:dipole}, 
\begin{equation}
S_{\text{reg}}=\frac{1}{2}\int_{z=\zeta}d^{d}xz^{-d}z\partial_{z}\left(\phi_{1}\phi_{2}\right).
\end{equation}
Using the mapping (\ref{eq:field_map}) and the propagators (\ref{eq:propagator}), we write 
\begin{equation}
\phi_{1}=\psi_{+}+\psi_{-},\quad\phi_{2}=\epsilon^{-1/2}\left(\psi_{+}-\psi_{-}\right),\quad\psi_{\pm}=\frac{1}{N_{\Delta_{\pm}}}\int d^{d}x^{\prime}\frac{z^{\Delta_{\pm}}u_{\pm}(x^{\prime})}{(z^{2}+\left|x-x^{\prime}\right|^{2})^{\Delta_{\pm}}},
\end{equation}
with $u_{\pm}$ given in (\ref{eq:sources_pm}).
Then $\phi_{1}\phi_{2}=\epsilon^{-1/2}(\psi_{+}^{2}-\psi_{-}^{2})$, so that all terms mixing $\Delta_{+}$ and $\Delta_{-}$ cancel.
From (\ref{eq: near_boundary_propagator}), $\psi_{\pm}=z^{d-\Delta_{\pm}}u_{\pm}+z^{\Delta_{\pm}}B_{\pm}+\cdots$ with $B_{\pm}(x)\equiv\frac{1}{N_{\Delta_{\pm}}}\int d^{d}x^{\prime}\frac{u_{\pm}(x^{\prime})}{\left|x-x^{\prime}\right|^{2\Delta_{\pm}}}$, and thus 
\begin{align}
S_{\text{reg}}= & \epsilon^{-1/2}\int_{z=\zeta}d^{d}x\left((d-\Delta_{+})z^{d-2\Delta_{+}}u_{+}^{2}-(d-\Delta_{-})z^{d-2\Delta_{-}}u_{-}^{2}\right)\nonumber \\
 & +\epsilon^{-1/2}\int d^{d}x\,d\left(u_{+}B_{+}-u_{-}B_{-}\right)+\cdots.
\end{align}
As in appendix~\ref{app:dipole}, the divergent terms are contact terms.
They are canceled by the counterterm (\ref{eq:counterterm}),
\begin{align}
S_{\text{ct}}= & \int_{z=\zeta}d^{d}x\sqrt{\gamma}\left(\frac{\phi_{1}^{2}+\epsilon\phi_{2}^{2}}{2\beta}-(d-\bar{\Delta})\phi_{1}\phi_{2}\right)\nonumber \\
= & -\epsilon^{-1/2}\int_{z=\zeta}d^{d}x\sqrt{\gamma}\left((d-\Delta_{+})\psi_{+}^{2}-(d-\Delta_{-})\psi_{-}^{2}\right),
\end{align}
which reduces to the counterterm in appendix~\ref{app:dipole} as $\epsilon\rightarrow0$ (derivative counterterms for $\mathrm{Re}(2\Delta_{\pm})-d\geq2$ are as in appendix~\ref{app:dipole}).
Since $\sqrt{\gamma}\psi_{\pm}^{2}=z^{d-2\Delta_{\pm}}u_{\pm}^{2}+2u_{\pm}B_{\pm}+\cdots$, the renormalized on-shell action, $S_{\text{ren}}\equiv\lim_{\zeta\rightarrow0}(S_{\text{reg}}+S_{\text{ct}})$, reads 
\begin{align}
S_{\text{ren}}= & \epsilon^{-1/2}\int d^{d}x\left((2\Delta_{+}-d)u_{+}B_{+}-(2\Delta_{-}-d)u_{-}B_{-}\right)\nonumber \\
= & \int d^{d}xd^{d}x^{\prime}\frac{\tilde{A}_{-}}{2}\frac{-\epsilon^{-1/2}\phi_{1(0)}(x)\phi_{1(0)}(x^{\prime})+2\phi_{1(0)}(x)\phi_{2(0)}(x^{\prime})-\epsilon^{1/2}\phi_{2(0)}(x)\phi_{2(0)}(x^{\prime})}{\left|x-x^{\prime}\right|^{2\Delta_{-}}}\nonumber \\
 & +\int d^{d}xd^{d}x^{\prime}\frac{\tilde{A}_{+}}{2}\frac{\epsilon^{-1/2}\phi_{1(0)}(x)\phi_{1(0)}(x^{\prime})+2\phi_{1(0)}(x)\phi_{2(0)}(x^{\prime})+\epsilon^{1/2}\phi_{2(0)}(x)\phi_{2(0)}(x^{\prime})}{\left|x-x^{\prime}\right|^{2\Delta_{+}}},
\end{align}
which is (\ref{eq:PT_regularizedPaction}), with $\tilde{A}_{\pm}\equiv A_{\pm}/N_{\Delta_{\pm}}$.
With the coupling (\ref{eq:GKPW}), the two-point functions at separated points are 
\begin{align}
\langle\mathscr{O}_{1}(x_{1})\mathscr{O}_{1}(x_{2})\rangle= & -\frac{\delta}{\delta\phi_{1(0)}(x_{1})}\frac{\delta}{\delta\phi_{1(0)}(x_{2})}S_{\text{ren}}|_{\phi_{1}(\partial\text{AdS})=\phi_{1(0)},\phi_{2}(\partial\text{AdS})=\phi_{2(0)}}\nonumber \\
= & \frac{1}{\sqrt{\epsilon}}\left(\frac{\tilde{A}_{-}}{\left|x_{1}-x_{2}\right|^{2\Delta_{-}}}-\frac{\tilde{A}_{+}}{\left|x_{1}-x_{2}\right|^{2\Delta_{+}}}\right),\\
\beta\langle\mathscr{O}_{1}(x_{1})\mathscr{O}_{2}(x_{2})\rangle= & -\frac{\delta}{\delta\phi_{1(0)}(x_{1})}\frac{\delta}{\delta\phi_{2(0)}(x_{2})}S_{\text{ren}}|_{\phi_{1}(\partial\text{AdS})=\phi_{1(0)},\phi_{2}(\partial\text{AdS})=\phi_{2(0)}}\nonumber \\
= & -\frac{\tilde{A}_{-}}{\left|x_{1}-x_{2}\right|^{2\Delta_{-}}}-\frac{\tilde{A}_{+}}{\left|x_{1}-x_{2}\right|^{2\Delta_{+}}},\\
\beta^{2}\langle\mathscr{O}_{2}(x_{1})\mathscr{O}_{2}(x_{2})\rangle= & -\frac{\delta}{\delta\phi_{2(0)}(x_{1})}\frac{\delta}{\delta\phi_{2(0)}(x_{2})}S_{\text{ren}}|_{\phi_{1}(\partial\text{AdS})=\phi_{1(0)},\phi_{2}(\partial\text{AdS})=\phi_{2(0)}}\nonumber \\
= & \sqrt{\epsilon}\left(\frac{\tilde{A}_{-}}{\left|x_{1}-x_{2}\right|^{2\Delta_{-}}}-\frac{\tilde{A}_{+}}{\left|x_{1}-x_{2}\right|^{2\Delta_{+}}}\right),
\end{align}
which are (\ref{eq:AdS_2pt_powerlaws}).
With $\left|x_{1}-x_{2}\right|^{-2\Delta_{\pm}}=\left|x_{1}-x_{2}\right|^{-2\bar{\Delta}}e^{\mp2\delta\ln\left|x_{1}-x_{2}\right|}$ and $\sqrt{\epsilon}=\beta\delta$, they give (\ref{eq:PT_AdS_2pt_1})--(\ref{eq:PT_AdS_2pt_3}).

\section{Two coupled hyperbolic circuits}\label{app:circuit}

Here, we recall the continuum limit of a single circuit \cite{Lenggenhager2022NC}, derive the voltage responses of the two coupled circuits of section~\ref{sec:circuit}, and describe the numerics.
The tessellation is drawn in the \Poincare ball of dimension $D=d+1$, with the metric $4|d\chi|^{2}/(1-|\chi|^{2})^{2}$ and $\chi\in\mathbb{R}^{D}$.
At the center, the metric is $4|d\chi|^{2}$, so that $\nabla_{\mu}\nabla^{\mu}=\partial_{\chi}^{2}/4$.
Consider a node $j$ at the center and its $q$ neighbors $k$ at $|\chi|=h$.
For a voltage that varies slowly over one edge, a Taylor expansion gives \cite{Lenggenhager2022NC}
\begin{equation}
\sum_{k}(V_{k}-V_{j})\simeq\frac{qh^{2}}{2D}\partial_{\chi}^{2}V=s\nabla_{\mu}\nabla^{\mu}V,\qquad s=\frac{2qh^{2}}{D},\label{eq:circuit_taylor}
\end{equation}
since the rotations about the center permute the neighbors, so that the first-order terms cancel and the second-order term is isotropic.
An isometry maps the tessellation onto itself and any node onto the center, so that eq.~(\ref{eq:circuit_taylor}) holds at every node whose neighbors all belong to the circuit.
For the $\{3,7\}$ tessellation, $D=2$, $h=[1-\sin^{2}(\pi/7)/\cos^{2}(\pi/3)]^{1/2}\approx0.497$, and $s=qh^{2}\approx1.729$ \cite{Lenggenhager2022NC,Dey2024PRL}.
For the $\{4,3,5\}$ honeycomb, $D=3$, $q=12$, and every edge has the geodesic length $\mathrm{arccosh}\,\varphi^{2}$, with $\varphi$ the golden ratio, so that $h=\tanh(\frac{1}{2}\mathrm{arccosh}\,\varphi^{2})$ and $s=8h^{2}\approx3.578$.

Two kinds of elements connect the nodes to ground \cite{Lenggenhager2022NC,Chen2023arXiv}.
The element at every node gives the term $y$ in eq.~(\ref{eq:circuit_node}): a capacitor gives $y>0$, and an inductor $L_{\mathrm{g}}$ gives $y=-1/(\omega^{2}L_{\mathrm{g}}C)<0$.
In addition, for each neighbor that an edge node lacks, a capacitor $C$ connects the edge node to ground, which imposes Dirichlet boundary conditions.

At every node, the capacitor $C_{b}$ and the INIC $C_{a}$ couple the voltages of the two circuits \cite{Helbig2020NP}.
A capacitor $2C_{a}$ from every node of the first circuit to ground makes their diagonal contributions equal in the two circuits, and we absorb these contributions into $y$.
When $C_{a}$ changes, the element to ground is adjusted so that $y$ stays fixed.
Equation~(\ref{eq:circuit_node}) for the two circuits then reads
\begin{equation}
\frac{1}{i\omega C}\begin{pmatrix}I_{1}\\
I_{2}
\end{pmatrix}=\begin{pmatrix}M & -C_{-}/C\\
-C_{+}/C & M
\end{pmatrix}\begin{pmatrix}V_{1}\\
V_{2}
\end{pmatrix},\qquad C_{\pm}\equiv C_{b}\pm C_{a},\label{eq:circuit_matrix}
\end{equation}
where $V_{a}$ and $I_{a}$ collect the voltages and currents of circuit $a$.
After division by $s$, the off-diagonal entries $-C_{\mp}/(sC)$ are the coefficients with which each circuit enters the equation of the other.
With $\phi_{1}=V_{1}$ and $\phi_{2}=-(sC/C_{+})V_{2}$, and the currents of the second circuit rescaled by the same factor, they become the coefficients $\epsilon$ and $1$ of eq.~(\ref{eq:PT_modified_EOM}).
The coupling matrix $-\frac{1}{C}\begin{pmatrix}0 & C_{-}\\ C_{+} & 0\end{pmatrix}$ has the eigenvalues $\mp\sqrt{C_{+}C_{-}}/C$, which shift $m^{2}=y/s$ to $m^{2}\mp\sqrt{C_{+}C_{-}}/(sC)$.
These are the squared masses $m^{2}\mp\sqrt{\epsilon}$ of the eigenmodes $\psi_{\mp}$ for $\epsilon=C_{+}C_{-}/(sC)^{2}$, which is eq.~(\ref{eq:circuit_eps}).
Since $C_{\pm}/C$ are numbers, the blocks of eq.~(\ref{eq:circuit_matrix}) commute, and the inverse reads
\begin{equation}
\begin{pmatrix}G_{11} & G_{12}\\
G_{21} & G_{22}
\end{pmatrix}=\left(M^{2}-s^{2}\epsilon\right)^{-1}\begin{pmatrix}M & C_{-}/C\\
C_{+}/C & M
\end{pmatrix}.\label{eq:circuit_inverse}
\end{equation}
Hence $G_{11}=G_{22}$ and $G_{12}/G_{21}=C_{-}/C_{+}=(C_{b}-C_{a})/(C_{b}+C_{a})$.
At the exceptional point, $C_{a}=C_{b}$ and $C_{-}=0$, so that $G_{11}=M^{-1}$, $G_{12}=0$, and $G_{21}=(2C_{b}/C)\,M^{-2}=-(2C_{b}/C)\,\partial M^{-1}/\partial y$.
Since the edge-to-edge entries of $M^{-1}$ decay as $\sigma^{-2\Delta(y)}$, the ratio $G_{21}/G_{11}$ is therefore linear in $\ln\sigma$, with the slope $4(C_{b}/C)\,d\Delta/dy=4(C_{b}/sC)\,d\Delta/dm^{2}$.

For $\epsilon\geq0$, that is, $C_{a}\leq C_{b}$, all off-diagonal entries of the matrix in eq.~(\ref{eq:circuit_matrix}) are nonpositive.
If, in addition, the smaller squared mass $m^{2}-\sqrt{\epsilon}$ lies above the Breitenlohner--Freedman bound of the lattice \cite{Basteiro2023PRL}, so that $M-s\sqrt{\epsilon}$ is positive definite, the inverse of this matrix has only nonnegative entries, so that every $G_{ab}$ is nonnegative.
For $\epsilon<0$, $G_{11}$ is the real part of the response of a single circuit at a complex squared mass, which oscillates in $\ln\sigma$.
A sign change of $G_{11}$ thus identifies the PT-broken phase.

We compute the voltages of ideal circuits.
For given injected currents, eq.~(\ref{eq:circuit_matrix}) is a system of $2N$ linear equations for the vector $(V_{1},V_{2})$ of the $2N$ voltage amplitudes, with $N$ the number of nodes per circuit, and its matrix is sparse.
With six rings (section~\ref{subsec:circuit_one}), the $\{3,7\}$ tessellation has $N=1625$ nodes, of which the $1008$ nodes of the outermost ring are edge nodes.
With seven shells, the three-dimensional counterpart of rings, the $\{4,3,5\}$ honeycomb has $N\approx3.6\times10^{6}$ nodes.
We inject a unit current at one edge node of one circuit at a time, for a set of edge nodes spread along the edge, and solve the resulting linear systems directly, by a sparse factorization of the matrix.
Since eq.~(\ref{eq:circuit_inverse}) holds on any lattice, the same voltages also follow from the two $N\times N$ matrices $M\mp s\sqrt{\epsilon}$ of the eigenmodes, which halves the size of the linear systems; we use this form for the $\{4,3,5\}$ honeycomb.
We then average $G_{ab}$ over pairs of a source node and an edge node with nearly equal $\ln\sigma$, which gives the $2\times2$ matrix $G(\sigma)$ as a function of $\sigma$.
The scaling dimensions follow from linear fits of the logarithms of the two eigenvalues of $G(\sigma)G(\sigma_{0})^{-1}$ against $\ln\sigma$, with the imaginary part of each logarithm unwrapped (followed continuously in $\ln\sigma$ instead of reduced to $(-\pi,\pi]$).
These fits use a range of $\ln\sigma$, chosen for each tessellation, in which the local exponent $-\frac{1}{2}d\ln|(M^{-1})_{jk}|/d\ln\sigma$ of a single circuit, for edge nodes $j$ and $k$, is approximately constant: away from the shortest separations, where the lattice spacing matters, and from the largest ones, where the finite size of the circuit matters.

On the $\{3,7\}$ tessellation with $C_{b}=C$, $\epsilon=0.2$ needs $C_{a}\approx0.63\,C$, and $\epsilon=-1$ needs $C_{a}\approx2.0\,C$.
The scaling dimensions of the circuits change little with the number of nodes, since the edge length of a regular tessellation is fixed in units of the AdS radius.
This edge length, $1.09$ for the $\{3,7\}$ tessellation and $1.62$ for the $\{4,3,5\}$ honeycomb, is not small compared with the unit AdS radius, so that the lattice corrections do not vanish as the number of nodes grows.
At the exceptional point, $\Delta$ deviates by $2\%$ and $3\%$ from the continuum value.
The splitting deviates more: at $\epsilon=-1$, $\mathrm{Im}\,\Delta_{\pm}$ is $12\%$ and $25\%$ smaller than in the continuum, and the slope of $G_{21}/G_{11}$ at the exceptional point is $13\%$ and $25\%$ smaller.
The slope of $G_{21}/G_{11}$ in figure~\ref{fig:circuit}(b) agrees within $1\%$ with $4(C_{b}/sC)\,d\Delta/dm^{2}$, with $d\Delta/dm^{2}$ from the splitting $\Delta_{+}-\Delta_{-}$ at $\epsilon=0.02$.
Since the number of nodes grows twice as fast with the geodesic distance from the center on the honeycomb, the same range of $\ln\sigma$ needs roughly the square of the number of nodes of the $\{3,7\}$ tessellation.

The two constants of the lines in figure~\ref{fig:circuit_measurement}(b) are the scaling dimension of the circuits at the exceptional point, which rescales the squared masses $m^{2}\pm\sqrt{\epsilon}$ by a common lattice factor, and the separation at which $G_{21}$ vanishes there.

\bibliographystyle{JHEP}
\bibliography{references}

\providecommand{\href}[2]{#2}\begingroup\raggedright\begin{thebibliography}{10}

\bibitem{GURARIE1993NPB}
V.~Gurarie, \emph{{Logarithmic operators in conformal field theory}},
  \href{https://doi.org/10.1016/0550-3213(93)90528-W}{\emph{Nucl. Phys. B}
  {\bfseries 410} (1993) 535}
  [\href{https://arxiv.org/abs/hep-th/9303160}{{\ttfamily hep-th/9303160}}].

\bibitem{Hogervorst2017JHEP}
M.~Hogervorst, M.~Paulos and A.~Vichi, \emph{{The ABC (in any D) of Logarithmic
  CFT}}, \href{https://doi.org/10.1007/JHEP10(2017)201}{\emph{JHEP} {\bfseries
  10} (2017) 201} [\href{https://arxiv.org/abs/1605.03959}{{\ttfamily
  1605.03959}}].

\bibitem{CAUX1996NPB}
J.-S.~Caux, I.I.~Kogan and A.M.~Tsvelik, \emph{{Logarithmic operators and
  hidden continuous symmetry in critical disordered models}},
  \href{https://doi.org/10.1016/0550-3213(96)00118-6}{\emph{Nucl. Phys. B}
  {\bfseries 466} (1996) 444}
  [\href{https://arxiv.org/abs/hep-th/9511134}{{\ttfamily hep-th/9511134}}].

\bibitem{Cardy2013JPA}
J.~Cardy, \emph{{Logarithmic conformal field theories as limits of ordinary
  CFTs and some physical applications}},
  \href{https://doi.org/10.1088/1751-8113/46/49/494001}{\emph{J. Phys. A}
  {\bfseries 46} (2013) 494001}
  [\href{https://arxiv.org/abs/1302.4279}{{\ttfamily 1302.4279}}].

\bibitem{Maldacena1999IJTP}
J.M.~Maldacena, \emph{{The Large $N$ limit of superconformal field theories and
  supergravity}}, \href{https://doi.org/10.4310/ATMP.1998.v2.n2.a1}{\emph{Adv.
  Theor. Math. Phys.} {\bfseries 2} (1998) 231}
  [\href{https://arxiv.org/abs/hep-th/9711200}{{\ttfamily hep-th/9711200}}].

\bibitem{witten1998ATMP}
E.~Witten, \emph{{Anti de Sitter space and holography}},
  \href{https://doi.org/10.4310/ATMP.1998.v2.n2.a2}{\emph{Adv. Theor. Math.
  Phys.} {\bfseries 2} (1998) 253}
  [\href{https://arxiv.org/abs/hep-th/9802150}{{\ttfamily hep-th/9802150}}].

\bibitem{GUBSER1998PLB}
S.S.~Gubser, I.R.~Klebanov and A.M.~Polyakov, \emph{{Gauge theory correlators
  from non-critical string theory}},
  \href{https://doi.org/10.1016/S0370-2693(98)00377-3}{\emph{Phys. Lett. B}
  {\bfseries 428} (1998) 105}
  [\href{https://arxiv.org/abs/hep-th/9802109}{{\ttfamily hep-th/9802109}}].

\bibitem{Grumiller2013JPA}
D.~Grumiller, W.~Riedler, J.~Rosseel and T.~Zojer, \emph{{Holographic
  applications of logarithmic conformal field theories}},
  \href{https://doi.org/10.1088/1751-8113/46/49/494002}{\emph{J. Phys. A}
  {\bfseries 46} (2013) 494002}
  [\href{https://arxiv.org/abs/1302.0280}{{\ttfamily 1302.0280}}].

\bibitem{GHEZELBASH1999IJMPA}
A.M.~Ghezelbash, M.~Khorrami and A.~Aghamohammadi, \emph{{Logarithmic conformal
  field theories and AdS correspondence}},
  \href{https://doi.org/10.1142/S0217751X99001287}{\emph{Int. J. Mod. Phys. A}
  {\bfseries 14} (1999) 2581}
  [\href{https://arxiv.org/abs/hep-th/9807034}{{\ttfamily hep-th/9807034}}].

\bibitem{KOGAN1999PLB}
I.I.~Kogan, \emph{{Singletons and logarithmic CFT in AdS / CFT
  correspondence}},
  \href{https://doi.org/10.1016/S0370-2693(99)00576-6}{\emph{Phys. Lett. B}
  {\bfseries 458} (1999) 66}
  [\href{https://arxiv.org/abs/hep-th/9903162}{{\ttfamily hep-th/9903162}}].

\bibitem{Flato1987CMP}
M.~Flato and C.~Fronsdal, \emph{{The Singleton Dipole}},
  \href{https://doi.org/10.1007/BF01212320}{\emph{Commun. Math. Phys.}
  {\bfseries 108} (1987) 469}.

\bibitem{Bergshoeff2012JHEP}
E.A.~Bergshoeff, S.~de~Haan, W.~Merbis, M.~Porrati and J.~Rosseel,
  \emph{{Unitary Truncations and Critical Gravity: a Toy Model}},
  \href{https://doi.org/10.1007/JHEP04(2012)134}{\emph{JHEP} {\bfseries 04}
  (2012) 134} [\href{https://arxiv.org/abs/1201.0449}{{\ttfamily 1201.0449}}].

\bibitem{Lu2011PRL}
H.~L{\"u} and C.N.~Pope, \emph{{Critical Gravity in Four Dimensions}},
  \href{https://doi.org/10.1103/PhysRevLett.106.181302}{\emph{Phys. Rev. Lett.}
  {\bfseries 106} (2011) 181302}
  [\href{https://arxiv.org/abs/1101.1971}{{\ttfamily 1101.1971}}].

\bibitem{Bergshoeff2011PRD}
E.A.~Bergshoeff, O.~Hohm, J.~Rosseel and P.K.~Townsend, \emph{{Modes of Log
  Gravity}}, \href{https://doi.org/10.1103/PhysRevD.83.104038}{\emph{Phys. Rev.
  D} {\bfseries 83} (2011) 104038}
  [\href{https://arxiv.org/abs/1102.4091}{{\ttfamily 1102.4091}}].

\bibitem{Johansson2012JHEP}
N.~Johansson, A.~Naseh and T.~Zojer, \emph{{Holographic two-point functions for
  4d log-gravity}}, \href{https://doi.org/10.1007/JHEP09(2012)114}{\emph{JHEP}
  {\bfseries 09} (2012) 114} [\href{https://arxiv.org/abs/1205.5804}{{\ttfamily
  1205.5804}}].

\bibitem{BenderMannheim2008PRD}
C.M.~Bender and P.D.~Mannheim, \emph{{Exactly solvable PT-symmetric Hamiltonian
  having no Hermitian counterpart}},
  \href{https://doi.org/10.1103/PhysRevD.78.025022}{\emph{Phys. Rev. D}
  {\bfseries 78} (2008) 025022}
  [\href{https://arxiv.org/abs/0804.4190}{{\ttfamily 0804.4190}}].

\bibitem{Mannheim2011GRG}
P.D.~Mannheim, \emph{{Comprehensive Solution to the Cosmological Constant,
  Zero-Point Energy, and Quantum Gravity Problems}},
  \href{https://doi.org/10.1007/s10714-010-1088-z}{\emph{Gen. Rel. Grav.}
  {\bfseries 43} (2011) 703} [\href{https://arxiv.org/abs/0909.0212}{{\ttfamily
  0909.0212}}].

\bibitem{Mannheim2023EPJP}
P.D.~Mannheim, \emph{{How to quantize gravity and how not to quantize
  gravity}}, \href{https://doi.org/10.1140/epjp/s13360-023-03826-4}{\emph{Eur.
  Phys. J. Plus} {\bfseries 138} (2023) 262}
  [\href{https://arxiv.org/abs/2209.15047}{{\ttfamily 2209.15047}}].

\bibitem{Fiorucci2024PRD}
A.~Fiorucci, D.~Grumiller and R.~Ruzziconi, \emph{{Logarithmic celestial
  conformal field theory}},
  \href{https://doi.org/10.1103/PhysRevD.109.L021902}{\emph{Phys. Rev. D}
  {\bfseries 109} (2024) L021902}
  [\href{https://arxiv.org/abs/2305.08913}{{\ttfamily 2305.08913}}].

\bibitem{Bissi2024JHEP}
A.~Bissi, L.~Donnay and B.~Valsesia, \emph{{Logarithmic doublets in CCFT}},
  \href{https://doi.org/10.1007/JHEP12(2024)031}{\emph{JHEP} {\bfseries 12}
  (2024) 031} [\href{https://arxiv.org/abs/2407.17123}{{\ttfamily
  2407.17123}}].

\bibitem{Bhardwaj2024JHEP}
R.~Bhardwaj and A.~Yelleshpur~Srikant, \emph{{Celestial soft currents at
  one-loop and their OPEs}},
  \href{https://doi.org/10.1007/JHEP07(2024)034}{\emph{JHEP} {\bfseries 07}
  (2024) 034} [\href{https://arxiv.org/abs/2403.10443}{{\ttfamily
  2403.10443}}].

\bibitem{Hao2022PRD}
P.-x.~Hao, W.~Song, X.~Xie and Y.~Zhong, \emph{{BMS-invariant free scalar
  model}}, \href{https://doi.org/10.1103/PhysRevD.105.125005}{\emph{Phys. Rev.
  D} {\bfseries 105} (2022) 125005}
  [\href{https://arxiv.org/abs/2111.04701}{{\ttfamily 2111.04701}}].

\bibitem{Bagchi2026EPJC}
A.~Bagchi, A.~Banerjee, P.~Dhivakar, S.~Mondal and A.~Shukla, \emph{{The
  Carrollian kaleidoscope}},
  \href{https://doi.org/10.1140/epjc/s10052-026-15437-1}{\emph{Eur. Phys. J. C}
  {\bfseries 86} (2026) 429}
  [\href{https://arxiv.org/abs/2506.16164}{{\ttfamily 2506.16164}}].

\bibitem{Vasseur2012JSTAT}
R.~Vasseur, J.L.~Jacobsen and H.~Saleur, \emph{{Logarithmic observables in
  critical percolation}},
  \href{https://doi.org/10.1088/1742-5468/2012/07/L07001}{\emph{J. Stat. Mech.}
  {\bfseries 1207} (2012) L07001}
  [\href{https://arxiv.org/abs/1206.2312}{{\ttfamily 1206.2312}}].

\bibitem{Gurarie1999NPB}
V.~Gurarie, \emph{{c-Theorem for Disordered Systems}},
  \href{https://doi.org/10.1016/S0550-3213(99)00013-9}{\emph{Nucl. Phys. B}
  {\bfseries 546} (1999) 765}
  [\href{https://arxiv.org/abs/cond-mat/9808063}{{\ttfamily
  cond-mat/9808063}}].

\bibitem{Skenderis2009JHEP}
K.~Skenderis, M.~Taylor and B.C.~van Rees, \emph{{Topologically Massive Gravity
  and the AdS/CFT Correspondence}},
  \href{https://doi.org/10.1088/1126-6708/2009/09/045}{\emph{JHEP} {\bfseries
  09} (2009) 045} [\href{https://arxiv.org/abs/0906.4926}{{\ttfamily
  0906.4926}}].

\bibitem{Grumiller2011JHEP}
D.~Grumiller, N.~Johansson and T.~Zojer, \emph{{Short-cut to new anomalies in
  gravity duals to logarithmic conformal field theories}},
  \href{https://doi.org/10.1007/JHEP01(2011)090}{\emph{JHEP} {\bfseries 01}
  (2011) 090} [\href{https://arxiv.org/abs/1010.4449}{{\ttfamily 1010.4449}}].

\bibitem{Ashida2020AP}
Y.~Ashida, Z.~Gong and M.~Ueda, \emph{{Non-Hermitian physics}},
  \href{https://doi.org/10.1080/00018732.2021.1876991}{\emph{Adv. Phys.}
  {\bfseries 69} (2020) 249}
  [\href{https://arxiv.org/abs/2006.01837}{{\ttfamily 2006.01837}}].

\bibitem{bender2023arXiv}
C.M.~Bender and D.W.~Hook, \emph{{$\mathcal{PT}$-symmetric quantum mechanics}},
  \href{https://doi.org/10.1103/RevModPhys.96.045002}{\emph{Rev. Mod. Phys.}
  {\bfseries 96} (2024) 045002}
  [\href{https://arxiv.org/abs/2312.17386}{{\ttfamily 2312.17386}}].

\bibitem{Jin2026arXiv}
Q.~Jin, K.~Ren, G.~Yang and R.~Yu, \emph{{Non-Hermitian Structure and
  Exceptional Points in Yang-Mills Theory from Analytic Continuation of
  $N_c$}},  \href{https://arxiv.org/abs/2603.19006}{{\ttfamily 2603.19006}}.

\bibitem{Chen2026arXiv}
J.~Chen, \emph{{When Conformal Interfaces Become Logarithmic}},
  \href{https://arxiv.org/abs/2610.23509}{{\ttfamily 2610.23509}}.

\bibitem{Kaplan2009PRD}
D.B.~Kaplan, J.-W.~Lee, D.T.~Son and M.A.~Stephanov, \emph{{Conformality
  Lost}}, \href{https://doi.org/10.1103/PhysRevD.80.125005}{\emph{Phys. Rev. D}
  {\bfseries 80} (2009) 125005}
  [\href{https://arxiv.org/abs/0905.4752}{{\ttfamily 0905.4752}}].

\bibitem{Gorbenko2018JHEP}
V.~Gorbenko, S.~Rychkov and B.~Zan, \emph{{Walking, Weak first-order
  transitions, and Complex CFTs}},
  \href{https://doi.org/10.1007/JHEP10(2018)108}{\emph{JHEP} {\bfseries 10}
  (2018) 108} [\href{https://arxiv.org/abs/1807.11512}{{\ttfamily
  1807.11512}}].

\bibitem{Faedo2020PRL}
A.F.~Faedo, C.~Hoyos, D.~Mateos and J.G.~Subils, \emph{{Holographic Complex
  Conformal Field Theories}},
  \href{https://doi.org/10.1103/PhysRevLett.124.161601}{\emph{Phys. Rev. Lett.}
  {\bfseries 124} (2020) 161601}
  [\href{https://arxiv.org/abs/1909.04008}{{\ttfamily 1909.04008}}].

\bibitem{Daniel_2020_SciPost}
D.~Are{\'a}n, K.~Landsteiner and I.~Salazar~Landea, \emph{{Non-hermitian
  holography}},
  \href{https://doi.org/10.21468/SciPostPhys.9.3.032}{\emph{SciPost Phys.}
  {\bfseries 9} (2020) 032} [\href{https://arxiv.org/abs/1912.06647}{{\ttfamily
  1912.06647}}].

\bibitem{Sergio_2023_SciPost}
S.~Morales-Tejera and K.~Landsteiner, \emph{{Non-Hermitian quantum quenches in
  holography}},
  \href{https://doi.org/10.21468/SciPostPhys.14.3.030}{\emph{SciPost Phys.}
  {\bfseries 14} (2023) 030}
  [\href{https://arxiv.org/abs/2203.02524}{{\ttfamily 2203.02524}}].

\bibitem{Meyer_2024_SciPost}
Z.-Y.~Xian, D.~Rodr{\'\i}guez~Fern{\'a}ndez, Z.~Chen, Y.~Liu and R.~Meyer,
  \emph{{Electric conductivity in non-Hermitian holography}},
  \href{https://doi.org/10.21468/SciPostPhys.16.1.004}{\emph{SciPost Phys.}
  {\bfseries 16} (2024) 004}
  [\href{https://arxiv.org/abs/2304.11183}{{\ttfamily 2304.11183}}].

\bibitem{Begines2026arXiv}
J.M.~Begines, A.F.~Faedo, C.~Hoyos and D.~Musso, \emph{{Supergravity flows,
  wormholes and their pseudo-Hermitian holographic duals}},
  \href{https://doi.org/10.1007/JHEP10(2026)108}{\emph{JHEP} {\bfseries 10}
  (2026) 108} [\href{https://arxiv.org/abs/2605.16168}{{\ttfamily
  2605.16168}}].

\bibitem{WangYang2026JHEP}
Y.~Wang and J.~Yang, \emph{{Robin holography in AdS and BTZ: double-trace RG
  flow and exceptional points}},
  \href{https://doi.org/10.1007/JHEP09(2026)090}{\emph{JHEP} {\bfseries 09}
  (2026) 090} [\href{https://arxiv.org/abs/2605.27641}{{\ttfamily
  2605.27641}}].

\bibitem{Long2026arXiv}
S.~Long, Y.~Shi, Z.~Xia, H.~Gong, Z.~Cao, Q.~Tan et~al., \emph{{Exceptional
  points and near-extremal boundary response in rotating BTZ black holes}},
  \href{https://arxiv.org/abs/2609.30612}{{\ttfamily 2609.30612}}.

\bibitem{Lenggenhager2022NC}
P.M.~Lenggenhager, A.~Stegmaier, L.K.~Upreti, T.~Hofmann, T.~Helbig,
  A.~Vollhardt et~al., \emph{{Simulating hyperbolic space on a circuit board}},
  \href{https://doi.org/10.1038/s41467-022-32042-4}{\emph{Nature Commun.}
  {\bfseries 13} (2022) 4373}
  [\href{https://arxiv.org/abs/2109.01148}{{\ttfamily 2109.01148}}].

\bibitem{Chen2023arXiv}
J.~Chen, F.~Chen, L.~Yang, Y.~Yang, L.~Chen, Z.~Chen et~al., \emph{{Testing
  holographic duality in hyperbolic lattices}},
  \href{https://arxiv.org/abs/2305.04862}{{\ttfamily 2305.04862}}.

\bibitem{Dey2024PRL}
S.~Dey, A.~Chen, P.~Basteiro, A.~Fritzsche, M.~Greiter, M.~Kaminski et~al.,
  \emph{{Simulating holographic conformal field theories on hyperbolic
  lattices}}, \href{https://doi.org/10.1103/PhysRevLett.133.061603}{\emph{Phys.
  Rev. Lett.} {\bfseries 133} (2024) 061603}
  [\href{https://arxiv.org/abs/2404.03062}{{\ttfamily 2404.03062}}].

\bibitem{Schindler2011PRA}
J.~Schindler, A.~Li, M.C.~Zheng, F.M.~Ellis and T.~Kottos, \emph{{Experimental
  study of active LRC circuits with PT symmetries}},
  \href{https://doi.org/10.1103/PhysRevA.84.040101}{\emph{Phys. Rev. A}
  {\bfseries 84} (2011) 040101}
  [\href{https://arxiv.org/abs/1109.2913}{{\ttfamily 1109.2913}}].

\bibitem{Helbig2020NP}
T.~Helbig, T.~Hofmann, S.~Imhof, M.~Abdelghany, T.~Kiessling, L.W.~Molenkamp
  et~al., \emph{{Generalized bulk-boundary correspondence in non-Hermitian
  topolectrical circuits}},
  \href{https://doi.org/10.1038/s41567-020-0922-9}{\emph{Nature Phys.}
  {\bfseries 16} (2020) 747}
  [\href{https://arxiv.org/abs/1907.11562}{{\ttfamily 1907.11562}}].

\bibitem{Stegmaier2021PRL}
A.~Stegmaier, S.~Imhof, T.~Helbig, T.~Hofmann, C.H.~Lee, M.~Kremer et~al.,
  \emph{{Topological defect engineering and PT symmetry in non-Hermitian
  electrical circuits}},
  \href{https://doi.org/10.1103/PhysRevLett.126.215302}{\emph{Phys. Rev. Lett.}
  {\bfseries 126} (2021) 215302}
  [\href{https://arxiv.org/abs/2011.14836}{{\ttfamily 2011.14836}}].

\bibitem{Pais1950PR}
A.~Pais and G.E.~Uhlenbeck, \emph{{On Field Theories with Non-Localized
  Action}}, \href{https://doi.org/10.1103/PhysRev.79.145}{\emph{Phys. Rev.}
  {\bfseries 79} (1950) 145}.

\bibitem{BenderMannheim2008PRL}
C.M.~Bender and P.D.~Mannheim, \emph{{No-ghost theorem for the fourth-order
  derivative Pais-Uhlenbeck oscillator model}},
  \href{https://doi.org/10.1103/PhysRevLett.100.110402}{\emph{Phys. Rev. Lett.}
  {\bfseries 100} (2008) 110402}
  [\href{https://arxiv.org/abs/0706.0207}{{\ttfamily 0706.0207}}].

\bibitem{Bergshoeff2011JHEP}
E.A.~Bergshoeff, S.~de~Haan, W.~Merbis and J.~Rosseel, \emph{{A
  Non-relativistic Logarithmic Conformal Field Theory from a Holographic Point
  of View}}, \href{https://doi.org/10.1007/JHEP09(2011)038}{\emph{JHEP}
  {\bfseries 09} (2011) 038} [\href{https://arxiv.org/abs/1106.6277}{{\ttfamily
  1106.6277}}].

\bibitem{Mannheim2018JPA}
P.D.~Mannheim, \emph{{Antilinearity Rather than Hermiticity as a Guiding
  Principle for Quantum Theory}},
  \href{https://doi.org/10.1088/1751-8121/aac035}{\emph{J. Phys. A} {\bfseries
  51} (2018) 315302} [\href{https://arxiv.org/abs/1512.04915}{{\ttfamily
  1512.04915}}].

\bibitem{Klebanov1999NPB}
I.R.~Klebanov and E.~Witten, \emph{{AdS / CFT correspondence and symmetry
  breaking}}, \href{https://doi.org/10.1016/S0550-3213(99)00387-9}{\emph{Nucl.
  Phys. B} {\bfseries 556} (1999) 89}
  [\href{https://arxiv.org/abs/hep-th/9905104}{{\ttfamily hep-th/9905104}}].

\bibitem{Faulkner2011JHEP}
T.~Faulkner, H.~Liu and M.~Rangamani, \emph{{Integrating out geometry:
  Holographic Wilsonian RG and the membrane paradigm}},
  \href{https://doi.org/10.1007/JHEP08(2011)051}{\emph{JHEP} {\bfseries 08}
  (2011) 051} [\href{https://arxiv.org/abs/1010.4036}{{\ttfamily 1010.4036}}].

\bibitem{MuckPRD1998}
W.~M{\"u}ck and K.S.~Viswanathan, \emph{{Conformal field theory correlators
  from classical scalar field theory on anti-de Sitter space}},
  \href{https://doi.org/10.1103/PhysRevD.58.041901}{\emph{Phys. Rev. D}
  {\bfseries 58} (1998) 041901}
  [\href{https://arxiv.org/abs/hep-th/9804035}{{\ttfamily hep-th/9804035}}].

\bibitem{Kostas2002CQG}
K.~Skenderis, \emph{{Lecture notes on holographic renormalization}},
  \href{https://doi.org/10.1088/0264-9381/19/22/306}{\emph{Class. Quant. Grav.}
  {\bfseries 19} (2002) 5849}
  [\href{https://arxiv.org/abs/hep-th/0209067}{{\ttfamily hep-th/0209067}}].

\bibitem{francesco2012conformal}
P.~Di~Francesco, P.~Mathieu and D.~S{\'e}n{\'e}chal, \emph{{Conformal Field
  Theory}}, Graduate Texts in Contemporary Physics, Springer-Verlag, New York
  (1997),
  \href{https://doi.org/10.1007/978-1-4612-2256-9}{10.1007/978-1-4612-2256-9}.

\bibitem{Wang2023arXivCLEP}
Z.~Wang and L.~He, \emph{{Emergent non-Hermitian conservation laws at
  exceptional points}},
  \href{https://doi.org/10.1103/PhysRevB.111.L100305}{\emph{Phys. Rev. B}
  {\bfseries 111} (2025) L100305}
  [\href{https://arxiv.org/abs/2408.01092}{{\ttfamily 2408.01092}}].

\bibitem{Mostafazadeh2002JMP}
A.~Mostafazadeh, \emph{{Pseudo-Hermiticity versus PT symmetry: The necessary
  condition for the reality of the spectrum of a non-Hermitian Hamiltonian}},
  \href{https://doi.org/10.1063/1.1418246}{\emph{J. Math. Phys.} {\bfseries 43}
  (2002) 205} [\href{https://arxiv.org/abs/math-ph/0107001}{{\ttfamily
  math-ph/0107001}}].

\bibitem{Balasubramanian1999PRD}
V.~Balasubramanian, P.~Kraus and A.E.~Lawrence, \emph{{Bulk versus boundary
  dynamics in anti-de Sitter space-time}},
  \href{https://doi.org/10.1103/PhysRevD.59.046003}{\emph{Phys. Rev. D}
  {\bfseries 59} (1999) 046003}
  [\href{https://arxiv.org/abs/hep-th/9805171}{{\ttfamily hep-th/9805171}}].

\bibitem{Fitzpatrick2011JHEP}
A.L.~Fitzpatrick, E.~Katz, D.~Poland and D.~Simmons-Duffin, \emph{{Effective
  Conformal Theory and the Flat-Space Limit of AdS}},
  \href{https://doi.org/10.1007/JHEP07(2011)023}{\emph{JHEP} {\bfseries 07}
  (2011) 023} [\href{https://arxiv.org/abs/1007.2412}{{\ttfamily 1007.2412}}].

\bibitem{Tang2024PRL}
Y.~Tang, H.~Ma, Q.~Tang, Y.-C.~He and W.~Zhu, \emph{{Reclaiming the Lost
  Conformality in a Non-Hermitian Quantum 5-State Potts Model}},
  \href{https://doi.org/10.1103/PhysRevLett.133.076504}{\emph{Phys. Rev. Lett.}
  {\bfseries 133} (2024) 076504}
  [\href{https://arxiv.org/abs/2403.00852}{{\ttfamily 2403.00852}}].

\bibitem{SonStarinets2002JHEP}
D.T.~Son and A.O.~Starinets, \emph{{Minkowski space correlators in AdS / CFT
  correspondence: Recipe and applications}},
  \href{https://doi.org/10.1088/1126-6708/2002/09/042}{\emph{JHEP} {\bfseries
  09} (2002) 042} [\href{https://arxiv.org/abs/hep-th/0205051}{{\ttfamily
  hep-th/0205051}}].

\bibitem{Birmingham2002PRL}
D.~Birmingham, I.~Sachs and S.N.~Solodukhin, \emph{{Conformal field theory
  interpretation of black hole quasinormal modes}},
  \href{https://doi.org/10.1103/PhysRevLett.88.151301}{\emph{Phys. Rev. Lett.}
  {\bfseries 88} (2002) 151301}
  [\href{https://arxiv.org/abs/hep-th/0112055}{{\ttfamily hep-th/0112055}}].

\bibitem{Peng2014NatPhys}
B.~Peng, {\c{S}}.K.~{\"O}zdemir, F.~Lei, F.~Monifi, M.~Gianfreda, G.L.~Long
  et~al., \emph{{Parity{\textendash}time-symmetric whispering-gallery
  microcavities}}, \href{https://doi.org/10.1038/nphys2927}{\emph{Nature Phys.}
  {\bfseries 10} (2014) 394} [\href{https://arxiv.org/abs/1308.4564}{{\ttfamily
  1308.4564}}].

\bibitem{Asaduzzaman2020PRD}
M.~Asaduzzaman, S.~Catterall, J.~Hubisz, R.~Nelson and J.~Unmuth-Yockey,
  \emph{{Holography on tessellations of hyperbolic space}},
  \href{https://doi.org/10.1103/PhysRevD.102.034511}{\emph{Phys. Rev. D}
  {\bfseries 102} (2020) 034511}
  [\href{https://arxiv.org/abs/2005.12726}{{\ttfamily 2005.12726}}].

\bibitem{Lee2018CommPhys}
C.H.~Lee, S.~Imhof, C.~Berger, F.~Bayer, J.~Brehm, L.W.~Molenkamp et~al.,
  \emph{{Topolectrical circuits}},
  \href{https://doi.org/10.1038/s42005-018-0035-2}{\emph{Commun. Phys.}
  {\bfseries 1} (2018) 39} [\href{https://arxiv.org/abs/1705.01077}{{\ttfamily
  1705.01077}}].

\bibitem{Bergshoeff2012PRD}
E.A.~Bergshoeff, S.~de~Haan, W.~Merbis, J.~Rosseel and T.~Zojer,
  \emph{{Three-dimensional tricritical gravity}},
  \href{https://doi.org/10.1103/PhysRevD.86.064037}{\emph{Phys. Rev. D}
  {\bfseries 86} (2012) 064037}
  [\href{https://arxiv.org/abs/1206.3089}{{\ttfamily 1206.3089}}].

\bibitem{Mannheim2012FP}
P.D.~Mannheim, \emph{{Making the Case for Conformal Gravity}},
  \href{https://doi.org/10.1007/s10701-011-9608-6}{\emph{Found. Phys.}
  {\bfseries 42} (2012) 388} [\href{https://arxiv.org/abs/1101.2186}{{\ttfamily
  1101.2186}}].

\bibitem{Basteiro2023PRL}
P.~Basteiro, F.~Dusel, J.~Erdmenger, D.~Herdt, H.~Hinrichsen, R.~Meyer et~al.,
  \emph{{Breitenlohner-Freedman bound on hyperbolic tilings}},
  \href{https://doi.org/10.1103/PhysRevLett.130.091604}{\emph{Phys. Rev. Lett.}
  {\bfseries 130} (2023) 091604}
  [\href{https://arxiv.org/abs/2205.05081}{{\ttfamily 2205.05081}}].

\end{thebibliography}\endgroup

\end{document}